\documentclass[11pt,a4paper]{article}
\pdfoutput=1
\usepackage{myjheppub}
\usepackage{latexsym,amsfonts,amsmath,amssymb,textgreek,upgreek}
\usepackage{amsthm}
\usepackage{mathtools}
\usepackage{bbm}
\usepackage{graphicx}
\usepackage{color}
\usepackage{amsfonts}
\usepackage{caption}
\usepackage{subcaption}
\usepackage[normalem]{ulem}
\usepackage{comment}
\usepackage{url}
\usepackage{slashed}
\usepackage{bm}
\usepackage{hhline}
\usepackage{physics}
\usepackage{booktabs}
\usepackage{siunitx}
\usepackage{tikz}
\usepackage{xcolor}
\usepackage{cases}

\usepackage{footmisc}

\usepackage{tikz}
\usetikzlibrary{arrows}

\newcommand{\CG}{\mathcal{G}}

\newcommand{\CN}{\mathcal{N}}

\newcommand{\IR}{\mathbb{R}}

\newcommand\be{\begin{equation}}
\newcommand\ee{\end{equation}}
\newcommand\bea{\begin{eqnarray}}
\newcommand\eea{\end{eqnarray}}

\renewcommand{\dd}{\mathrm{d}}

\renewcommand{\=}{\;= \;}

\newcommand{\s}{\sigma}

\newcommand{\ve}{\varepsilon}

\renewcommand{\v}{\varphi}

\newcommand{\ndt}{\noindent}

\newcommand{\tO}{\text{O}}

\def\spp{\;\;\;,\;\;\;}

\newcommand{\Qext}{Q_\star}
\newcommand{\Mext}{M_\star}
\newcommand{\rext}{r_\star}

\newcommand{\muext}{\mu}

\preprint{\\
	\hspace*{12cm} CCTP-2025-16\\
	\hspace*{12.1cm} ITCP-2025/16\\ }

\title{Holographic shear correlators at low temperatures, and quantum~$\eta/s$}

\author{Alexandros Kanargias${}^1$, Elias Kiritsis${}^{1,2}$, Sameer Murthy${}^3$,
Olga Papadoulaki${}^4$, Achilleas P. Porfyriadis${}^1$}

\affiliation{${}^1$ \href{https://hep.physics.uoc.gr/}{Crete Center for Theoretical Physics},
Institute of Theoretical and Computational Physics,\\
Department of Physics, University of Crete, 70013 Heraklion, Greece}
\affiliation{${}^2$ \href{http://www.apc.univ-paris7.fr}
{APC, Universit\'e Paris 7}, CNRS/IN2P3, CEA/IRFU, Obs. de Paris, Sorbonne Paris Cit\'e, B\^atiment Condorcet,
F-75205, Paris Cedex 13, France (UMR du CNRS 7164).}
\affiliation{${}^3$ \href{https://www.kcl.ac.uk/mathematics}{Department of Mathematics, King's College London},
The Strand, London WC2R 2LS, UK}
\affiliation{${}^4$ CPHT, CNRS, \'{E}cole polytechnique, Institut Polytechnique de Paris,  91120 Palaiseau, France}

\emailAdd{kanargias@physics.uoc.gr,
sameer.murthy@kcl.ac.uk,
olga.papadoulaki@polytechnique.edu, porfyriadis@physics.uoc.gr}

\abstract{The strongly-coupled 3-dimensional theory, holographically dual to
black branes at fixed chemical potential $\muext$ and temperature $T \ll \mu$ is considered
in AdS$_4$ Einstein-Maxwell theory.
The retarded Green's functions at frequency~$\omega$ is calculated using holography
in the regime~$\omega, T \ll \muext$ but otherwise arbitrary.
When the transverse space has  finite volume, there is a non-zero energy scale~$E_\text{gap}$,
scaling as~$1/\mu$ for large~$\mu$, below which quantum-gravitational corrections
due to the fluctuations of the nearly-gapless Schwarzian modes become important.
Such  corrections to the retarded Green's function are calculated at different relative
values of $\omega$, $T$, and~$E_\text{gap}$.
The~$\omega \to 0$ limit is used to define the shear viscosity~$\eta$.
As the temperature is lowered below~$\mu$,
quantum corrections are found to increase the value of~$\eta$ with respect to its semiclassical value.
The quantum-corrected result for~$\eta$ diverges as~$\sqrt{E_\text{gap}/T}$ at $T \ll E_\text{gap}$,
in accord with  corresponding results for the absorption cross section.
The quantum result for the ratio~$\eta/s$, where~$s$ is the entropy density,
dips below the semiclassical limit of~$1/4\pi$ when~$E_\text{gap} \ll T \ll \mu$,
then turns back to increase towards lower temperatures, and finally diverges at temperatures much below~$E_\text{gap}$.
}

\begin{document}

\maketitle

\section{Introduction and summary of results}

It is well-known that extremal black hole solutions of general relativity, coupled to matter fields in $n+2$ dimensions,
universally feature a region near their horizon in which the geometry is AdS$_2\times X_n$, ~\cite{Kunduri:2007vf}.
The manifold $X_n$  depends on the details of the theory and of the solution:
 $X_n=S^n$ for spherically symmetric black holes, while $X_n=\IR^n$, or~$T^n$, for non-compact,
or periodically-identified, black branes, respectively.
At small temperatures, when the black hole is nearly extremal, the solution develops a throat region
connecting the asymptotic region to a nearly AdS$_2$ geometry.
The throat becomes longer as one reduces the temperature and,
in the~$T\to 0$ limit, the throat becomes infinitely long,
the geometry develops an~AdS$_2$ factor as above, and the near-horizon
region becomes a novel arena for IR physics.

In recent years, it has become clear that the true description of near-extremal black holes
in the full quantum theory can be very different from the above semiclassical picture,
due to large quantum fluctuations of a nearly-gapless mode
in the throat region~\cite{Almheiri:2014cka,kitaev,Sachdev:2015efa,
Almheiri:2016fws,Maldacena:2016hyu,Engelsoy:2016xyb,Maldacena:2016upp,Jensen:2016pah,Mertens:2022irh}.
In particular, we have learned that (in the absence of supersymmetry)
there is no energy gap separating the states in the continuum
of energies above extremality from the extremal energy (ground)states, and that
the quantum entropy diminishes drastically as~$T \to 0$.

This has prompted a revisiting of many semi-classical notions at low temperatures
such as the density of states in different types of black
holes~\cite{Iliesiu:2020qvm,Heydeman:2020hhw,Iliesiu:2022onk,Iliesiu:2022kny},
scattering of waves from the black hole~\cite{Emparan:2025sao, Biggs:2025nzs, Emparan:2025qqf, Betzios:2025sct},
and Hawking radiation~\cite{Brown:2024ajk}.
In this paper we revisit, with a similar lens, correlators in the holographic theory
dual to near-extremal black branes in the quantum regime.
In the rest of this introduction, we introduce and summarize the main ideas and results of this paper.

\vskip 0.2cm

\ndt {\bf Quantum near-extremal black holes}

\vskip 0.1cm

\ndt Near-extremal black holes have many special features that indicate
that their thermodynamic interpretation has subtleties related to quantum effects~\cite{Preskill:1991tb}.
For a long time, the precise origin of these quantum effects was not clear.
An important observation emerged around ten years ago, when the study of a one-dimensional quantum theory
with quenched disorder, known as the SYK Model~\cite{kitaev,Sachdev:1992fk,Maldacena:2016hyu}, led to a conjectured
IR holographic duality with (two-dimensional)
JT gravity~\cite{Maldacena:2016upp,Jensen:2016pah,Engelsoy:2016xyb,Mertens:2022irh}.
JT gravity has a nearly AdS$_2$ solution, which is the same two-dimensional factor that appears in the
near-horizon region of near-extremal black holes.

\smallskip

A crucial insight obtained by the SYK/JT-gravity duality is the fact that,
even after taking the large-$N$ or semiclassical limit, there is a set of modes in the theory
whose quantum fluctuations are unsuppressed in the extremal (scale invariant) limit
associated to AdS$_2$.
Upon turning on a small temperature, the quantum part of JT gravity can be described in terms of a
boundary ``Schwarzian'' theory which provides the dynamics to
the pseudo-Goldstone mode arising from the breaking of the reparameterization invariance
of the one-dimensional boundary of Euclidean~AdS$_2$.
Importantly, the Schwarzian theory can be treated quantum-mechanically, although the rest of gravity is classical.
Its quantum effects do not decouple as the Planck mass scale becomes arbitrarily
large.\footnote{The fact that one-loop quantum gravitational effects can qualitatively alter
tree-level results in holography was already indicated in \cite{Anninos:2010sq}.}

The action of the Schwarzian theory is non-linear and contains four derivatives.
The partition function of the Schwarzian theory is one-loop exact~\cite{Stanford:2017thb}.
More generally, the theory is exactly solvable, and the correlators can be calculated by a variety of
methods (see the review~\cite{Mertens:2022irh}).
The same Schwarzian mode appears as a collective mode of the dual SYK model.
The part of the SYK/JT duality described by the Schwarzian is
believed to capture many of the universal aspects of the quantum dynamics of the nearly AdS$_2$ geometry
in the near-horizon region of near-extremal black holes.

\smallskip

An explanation of the Schwarzian modes and the consequent universality
can be reached from a slightly different point of view based on the~$n+2$-dimensional
low-energy effective theory~\cite{Iliesiu:2020qvm,Iliesiu:2022onk},
making contact with the program of calculating quantum corrections to extremal black hole entropy~\cite{Sen:2008yk}.
The semiclassical solution with the near-horizon AdS$_2$ region carries an enormous amount of entropy given by
the Bekenstein-Hawking formula~$S_0=A/4G$ of the extremal black hole. However, this large entropy leads to
puzzles~\cite{Preskill:1991tb} including a violation of the third law of thermodynamics. For asymptotically AdS extremal black holes,
which are dual to a holographic  QFT at finite density, this implies a similar violation in the dual quantum-field theoretic
systems.

\smallskip

The quantum entropy program of~\cite{Sen:2008yk} proposed that the entropy of extremal black holes
in the quantum theory is given by a functional integral of the gravitational theory over the
near-horizon~AdS$_2 \times X_n$ region. While this has been very successful in calculating corrections
suppressed by large charges and matching them to microscopic string theory
calculations~\cite{Banerjee:2010qc,Sen:2012kpz,Dabholkar:2011ec,Murthy:2015yfa},
there is a subtlety in the functional integral that needs to be considered carefully. The point is that
the AdS$_2 \times X_n$ region admits a set of zero modes to the Laplacian of any gauge field,
including the metric~\cite{Camporesi:1995fb}. The volume of this space of zero modes multiplies
the path integral over the rest of the gravitational fields, and needs to be considered more carefully.

This can be done by regulating the zero modes by a small temperature, and the resulting~$(n+2)$-dimensional
Einstein-matter action of these nearly-zero modes turns out to be precisely the Schwarzian action~\cite{Iliesiu:2020qvm,Iliesiu:2022onk},
with inverse coupling denoted by~$T/E_\text{gap}$.
Here~$E_\text{gap}$ is a new energy scale that arises in the quantum theory, at which the modes
which would become gapless at extremality become strongly coupled. It typically scales
as an inverse power of the entropy of the black hole and, therefore, is extremely small for a large black hole.

\smallskip

For the four-dimensional black brane wrapped on a 2-torus that we discuss in this paper,
$E_\text{gap}$ scales as~$(\mu V_2)^{-1}$, where~$\mu$ is the chemical potential of the theory
and~$V_2$ is the volume of the torus, cf.~\eqref{eq:defEgap}.
The quantum effects of the Schwarzian, substantially change the thermodynamics at~$T\ll E_\text{gap}$
and resolve some of the puzzles associated to near-extremal black holes.
In particular, the regulated volume of the space of Schwarzian modes is proportional to~$T^\frac32$,
and, the quantum corrected entropy correspondingly diminishes as $T\to 0$ in accord with the third law of
thermodynamics and with the expectations of \cite{Preskill:1991tb}.\footnote{The Schwarzian results
cannot be trusted at scales $T\sim \tO(e^{-S_0})$ or below, because of potential non-perturbative corrections
e.g. due to other saddles that may appear at that scale. Here, and below, when we refer to~$T=0$,
we mean an exponentially small temperature cutoff above which the Schwarzian results can be trusted to very good accuracy.}
The result of the quantum corrections is to push the $e^{S_0}$ zero-energy states to a continuum
above $E=0$ and the number of states populating the low-energy region is reduced with respect
to the original number of zero-energy states.\footnote{
In a dual quantum theory, like the SYK Model, such low-energy spectrum is dense, but discrete.
However, the discreteness is not expected to arise at any order in perturbation theory in the gravitational variables,
and will only arise at non-perturbatively small scales.}
\footnote{There is an exception to these conclusions in the case of supersymmetric near-extremal black holes.
In this case, the supersymmetric quantum states form an independent quantum system decoupled from the
full theory by a gap of the order of $E_\text{gap}$ \cite{Iliesiu:2022kny}, so as to recover an integer number of
states~\cite{Dabholkar:2010uh,Dabholkar:2011ec,Dabholkar:2014ema,Iliesiu:2022kny},
therefore alleviating differently the problems pointed out in \cite{Preskill:1991tb}.}

\vskip 0.2cm

\ndt {\bf Near-extremal dynamics in holography}

\vskip 0.1cm

\ndt

There are two aspects of the near-extremal black-hole story that complement the discussion above,
and provide novel physical problems. The first involves the low-energy dynamics associated with
near-extremal asymptotically AdS black holes. According to the AdS/CFT correspondence, such
dynamics should be mapped to various forms of hydrodynamics of the dual QFT
when~$\omega, {q^2} \ll T$, where~$\omega$ and~$\vec{q}$ are the frequency and momenta,
respectively, of the collective excitations, ~\cite{Shiraz,erd,Banerjee:2008th}, and $q\equiv |\vec q|$.

For the particular example of near-extremal Reissner-Nordstr\"{o}m (RN) black holes, the dual CFT
is at finite chemical potential $\mu$, and near extremality implies that $T\ll \mu$.
In~\cite{DP}, motivated by the results of \cite{EJL1, EJL3},  the low-energy dynamics of planar
RN-AdS$_4$ black holes was investigated. It was shown that  at any temperature, the low-energy
(with respect to $\mu$) collective excitations of the transverse components of the energy-momentum
tensor and the global $U(1)$ current in the dual QFT are simply those of hydrodynamics.
This suggested that hydrodynamics is applicable even when $T\ll \omega, q\ll \mu$.

This analysis was subsequently refined in \cite{Blaise,Blaise2}, where the structure of the poles of the energy-momentum
tensor and current correlators was computed numerically and followed around as the momentum $q$ was varied.
In the near-extremal limit there are three families of poles:

(a) Massless  ``hydrodynamic'' poles,

(b) AdS$_2$ poles whose positions on the  complex $\omega$ plane, (``masses"), are controlled by the temperature $T$, and

(c) AdS$_4$ poles whose positions (``masses") are controlled by $\mu\gg T$.

In the standard hydrodynamic limit, $\omega,q\ll T$, the hydrodynamic poles are well below the AdS$_2$ poles and the standard picture holds.

This breaks down when the massless poles approach the AdS$_2$ poles.
In the unusual regime, $T\ll \omega,q\ll \mu$, the hydrodynamic poles move inside the ``sea" of AdS$_2$ poles. Although, standard hydrodynamics is not expected to hold here,
it was shown in \cite{Blaise} that the identity of the hydrodynamic poles is well defined even in that case.

A derivation of the equations of low-energy dynamics  in the non-standard regime  $T\ll \omega, q\ll \mu$ was presented in \cite{Trivedi} using the holographic techniques of \cite{Shiraz}. They have shown that up to first order, the equations are those of first order hydrodynamics, with the standard first order transport coefficients. However in higher order, the effect of the AdS$_2$ poles (that become a branch cut in the limit $T\ll \omega, q\ll \mu$) is to introduce non-local in time behavior.

A similar effect was seen in  \cite{Blaise2} where the current-current correlator was computed  in the transition regime $\omega,q\sim T$. The effects of the AdS$_2$ branch-cut appeared in the logarithmic behavior in $\omega$ of the correlator. Moreover, the complex behavior of hydrodynamic poles as they are moving closer to the AdS$_2$ poles was revealed.

A numerical evaluation of the current-current correlators in the near-extremal RN background was presented in \cite{preau}. It was verified that the correlators matched the first-order hydrodynamic expression in the whole range $\omega,q\ll \mu$ providing a credible check of the approximate methods mentioned earlier. Moreover, this comparison indicated that the correlator matches its hydrodynamic approximation even outside the
$\omega,q <\mu$ regime if one is sufficiently close to the hydrodynamic pole.

It is interesting to note that for a near-extremal black brane, where the transverse space is a torus with radii $R_i$, the values of the spatial momentum $\vec q$ are quantized. For the four-dimensional case we analyze in this paper, the torus is two-dimensional\footnote{We consider here an orthogonal torus for simplicity.}  and
\be
\vec q=\left({m_1\over R_1},{m_2\over R_2}\right)\spp m_{1,2}\in \mathbb{Z}
\label{KK}\ee
Assuming that $|\vec q|<\mu$ we obtain that $\mu R_{1,2}$ are constrained\footnote{This is the case when the two radii $R_1$ and $R_2$ are of the same order.}.
We have roughly two asymptotic regimes:

$\bullet$  $\mu^2 V_2 \gg 1$ where $V_2=R_1R_2$ is proportional to the area of the torus. In this case, there is a large number of momentum KK modes that belong to the extended  hydrodynamic regime, $|\vec q|\ll \mu$, making a continuum in the spatial momenta, and the analogue of the extended hydrodynamic equations are (1+2)-dimensional, involving $\omega, \vec q$.

$\bullet$  $\mu^2 V_2 \ll 1$. In this case, only $\vec q=0$ belongs to the hydro regime and the analogue of extended hydrodynamics is (1+0)-dimensional
with no spatial dependence.
{Hence,}
this type of dynamics is trivial, stating the Energy and Charge are constant. It may become non-trivial if there is external forcing.

Overall, there are three dimensionless parameters that control effects in (extended) hydrodynamics. ${T\over \mu}\ll 1$, $\mu^2 V_2$ that controls the dimensionality of the hydrodynamic equations, and ${\cal N}\gg 1$, defined in (\ref{nu}), which is a dimensionless number that is proportional to the number of degrees of the dual CFT. This is a number that must be large in order for the holographic correspondence to be applicable at the semiclassical regime.

In addition, as we mentioned above, and discuss in more detail in the next section, the scale of quantum effects is
set by~$E_\text{gap}$ defined in (\ref{eq:defEgap}), or, equivalently, its inverse~$C$ given in (\ref{c}).
The corresponding dimensionless parameter
\be
CT \= {T\over E_{gap}} \; \sim \; {\cal N}\cdot \mu^2 V_2\cdot {T\over \mu}
\label{ct}
\ee
controls the Schwarzian quantum effects.

There are two regimes involving these three dimensionless numbers.

\vspace{-0.2cm}

\begin{enumerate}

\item[A.]  ${T\over \mu}\ll 1 \ll {\cal N}, \mu^2 V_2$. Since $\mu^2~V_2\gg 1$, the dynamics is three-dimensional in this case.
The parameter (\ref{ct}) can be either much smaller or much larger than one in this regime.
For fixed and large $\mu^2V_2$ and ${\cal N}$, we can choose ${T\over \mu}$ sufficiently small
so that $CT<1$ and  we shall be in the quantum regime.

\item[B.]  ${T\over \mu}, \mu^2 V_2\ll 1\ll {\cal N}$. Since $\mu^2V_2\ll 1$, the dynamics is one-dimensional in this case.
The parameter (\ref{ct})
 can be either much smaller or much larger than one in this regime. Here, for fixed and large ${\cal N}$,
 we can choose ${T\over \mu}$ and/or $\mu^2V_2$ sufficiently small so that $CT<1$ and  we shall be in the quantum regime.

\end{enumerate}

So far we were in the extended hydrodynamic regime with $T\ll \mu$. If, on the other hand, 
we consider the conventional hydrodynamic regime $\omega \ll T$, $q\ll T$, as was shown 
in~ \cite{Blaise,Blaise2}, it extends to $\omega \ll T$, $q\ll \sqrt{\mu T}$. For $q\sim  \sqrt{\mu T}$ 
the first collision with a AdS$_2$ pole happens.
Demanding that $q<\sqrt{\mu T}$, in view of (\ref{KK}), it implies that $(\mu^2 V_2)\cdot  {T\over \mu} \gg 1$ for the hydrodynamics to be three-dimensional. Otherwise the hydrodynamics degenerates to one-dimensional.
If on the other hand, $(\mu^2 V_2)\cdot  {T\over \mu} \gg 1$, then by (\ref{ct}), $CT>1$ and we are in the semiclassical regime, \cite{Gouteraux:2025exs}.

The conclusion is that for the conventional regime of three-dimensional hydrodynamics, we should have $CT>1$ and the quantum effects will be expected to give subleading corrections.
But in the extended hydro regime with $\omega,q\ll \mu$, there is room for quantum effects being important\footnote{ In the case of a very anisotropic torus eg. $R_1\gg R_2$, there are changes in the conclusions discussed in the text. In particular, for finite-volume but highly-anisotropic tori (ie. long strips) one can obtain (1+1)-dimensional hydrodynamics as an intermediate option.    }   .

\medskip

We now turn to the question of how the Schwarzian quantum corrections
affect the low-energy dynamics in the near-extremal black brane and, by holography, the dual ``cold" but ``dense" CFT.

\medskip

\ndt {\bf Quantum effects at low temperatures}

\vskip 0.1cm

In this paper, we focus on the quantum effects  on the retarded Green's function of the
stress tensor of the boundary theory. From the holographic point of view, this is achieved by
studying the behaviour of  the transverse and traceless bulk gravitational
mode~\cite{Policastro:2001yc,Policastro:2002se}. This reduces to the study of a massless neutral
scalar field propagating in the background of the asymptotically $AdS_4$ Reisner-Nordstr\"{o}m
black brane.
In fact, such a study was initiated in the paper~\cite{Daguerre:2023cyx}, where the two-point function
of an operator of dimension~$\Delta$ in the Schwarzian theory was discussed in this context.
In the present paper we take this forward by focussing on the operator with~$\Delta=1$, which
corresponds to the~$q=0$ mode of the four-dimensional field.

In the semiclassical case, the IR~$\omega=0$ limit of the associated retarded correlators evaluated at $q=0$
usually leads to the notion of transport coefficients---which then appear as certain coefficients in the
leading and higher order terms of the hydrodynamic equations.
In particular, the low-frequency, zero-momentum limit of the retarded Green's function of the transverse and traceless bulk gravitational mode gives the shear viscosity~$\eta$.
Here, we do not develop the higher-order equations, and take the IR limit as a definition of~$\eta$.
With this definition, our explicit results for the quantum correlators lead to formulas for the
quantum-corrected~$\eta$ as well as the quantum corrections to the semiclassical formula
of the ratio of the shear viscosity to the entropy density~$\eta/s$.

In Section \ref{sec:scbranes} we present our set-up and review the asymptotically $AdS_4$ Reisner-Nordstr\"{o}m
black brane and its near-extremal limit. We point out how the scale $E_{gap}$---that governs the fluctuations of the nearly
gapless modes of the theory---appears in this limit.
In Section \ref{sec:SCGreen} we review the holographic computation of $\eta/s$ in the semiclassical regime.

In Section \ref{sec:quGreen} we compute, in the bulk theory, the real time Wightman two-point function,
and the imaginary part of the retarded Green's function for a real operator
with conformal dimension $\Delta=1$, incorporating the Schwarzian corrections.

We  consider the ordering of the
following dimensionless variables, ${\omega\over T}$,   ${T\over E_{gap}}$ and ${T^2\over E^2_{gap}}$,
always assuming that $T,\omega, E_{gap} \ll \mu$.

\medskip

\ndt
Broadly, the regions are classified into two regimes: Regime~I for~$\omega \ll T$, and  Regime~II for~$T \ll \omega$.
Each regime is further divided into three regions according to the relative placement of the dimensionless
variables.
We give a brief summary below, and refer the reader to Section~\ref{sec:quGreen} and Appendix~\ref{app:intcalcs}
for more details.
In all regimes, the effects proportional to~$T/\mu$ and~$\omega/\mu$ are suppressed in all calculations.
In addition, the Green's functions can be approximated by different formulas
depending on the small parameters in the respective regimes in the Schwarzian theory.

\bigskip

\ndt \textbf{Regime I ($\omega \ll T$)\label{I}}

\vspace{-0.1cm}

\begin{enumerate}

\item[1.]
$E_{gap}\ll\omega\ll T$. In this case, $T,\omega \gg E_{gap}$, and therefore, both the background
and the dynamical modes are semiclassical. We therefore  expect a semiclassical description of the dynamics.

\item[2.]
$\omega\ll E_{gap}\ll T$. In this case $T\gg E_{gap}$, therefore, the background is in  the semiclassical  regime.
We therefore expect a semiclassical description of the dynamics. Although the dynamical modes
have $\omega\ll E_{gap}$ we do not expect quantum modifications of their behaviour.
Since ${\omega\ll T}$, we expect  to have the standard hydrodynamic equations without logarithmic corrections
in $\omega$ due to the AdS$_2$ poles.

In both regions~1 and~2, we find that the leading order result for the imaginary part of the retarded IR Green's function is
\be
\text{Im} \, \CG^{\Delta=1}_{_R} (\omega)  \= \omega + \dots \,,
\ee
in agreement with the semiclassical calculation.
The details, including the nature of the corrections to the semiclassical formula are given in
Equation~\eqref{eq:omlTquant2a} and the preceding discussion.

\item[3.]
$\omega\ll \frac{T^2}{E_\text{gap}} \ll E_{gap}$. In this case $ {T\ll E_{gap}}$  and $\omega\ll E_{gap}$,
and therefore, both the background and the dynamical modes are in the quantum regime.
It is not clear if the low-energy modes have an effective semiclassical
description in this regime. This will depend crucially on how big the quantum fluctuations are
compared to average values. If they are small, then a Langevin-type description may be possible with
quantum fluctuations added as a small correction around the modified classical evolution.
We find
\be \label{eq:smallomintro}
\text{Im} \, \CG^{\Delta=1}_{_R} (\omega)  \=  \dfrac{\sqrt{2} \, \omega}{\sqrt{\pi^3 C T}} + \dots  \,.
\ee
In particular the relevant IR operator still have $\Delta=1$ but its normalization is now $T$-dependent. The details, including the nature of the corrections to the above formula are given in Equation~\eqref{eq:omlTquant2b}
and the preceding discussion.

\medskip

The result for the imaginary part of the retarded IR Green's function for the regions~1--3
can be summarized in the single formula~\eqref{eq:omlTquant2}, which can be approximated
by~\eqref{eq:omlTquant2a} and~\eqref{eq:omlTquant2b} in the respective regimes mentioned above.

\end{enumerate}

\medskip

\ndt \textbf{Regime II ($T \ll \omega$)}

\vspace{-0.2cm}

\begin{enumerate}

\item[4.]

$E_{gap}\ll \frac{T^2}{E_\text{gap}} \ll \omega$.
In this case $ {E_{gap} \ll T}$  and $E_{gap}\ll \omega$,
and therefore, both the background and the fluctuations lie in the semiclassical regime.
We therefore expect a semiclassical description of the dynamics.  Since $\omega\gg T$ we expect
to have the modified hydrodynamic equations with the logarithmic corrections in $\omega$ originating from
the influence of the condensed AdS$_2$ poles (that behave as a branch cut in this regime).

\item[5.]
$T\ll E_{gap}\ll \omega$.  In this case $ {T\ll E_{gap}}$  and the background system is in the quantum regime. Since $\omega\gg E_{gap}$ the dynamical modes are well above the energy cutoff, $E_{gap}$, that controls the low-energy quantum effects. It is therefore plausible that although the thermodynamics is corrected by the quantum effects, the dynamical evolution is semiclassical and in the non-standard hydrodynamic regime with potentially different,  however, transport coefficients.
    Given that the AdS$_2$ poles lie both below and above $E_{gap}$ in this case, we might expect that those that existed well above $E_{gap}$ escape unscathed from the Schwarzian quantum effects. They are expected to influence the hydrodynamics probably via logarithmic corrections.

In both regions~4 and~5, we find that the leading order result for the imaginary
part of the retarded IR Green's function is also given by
\be
\text{Im} \, \CG^{\Delta=1}_{_R} (\omega)  \= \omega + \dots \,,
\ee
in agreement with the semiclassical calculation.
The details, including the nature of the corrections to the semiclassical formula are given in
Equation~\eqref{eq:Tlomquanta} and the preceding discussion.

\item[6.]
$T\ll \omega\ll E_{gap}$.  In this case, $ {T\ll E_{gap}}$  and $\omega\ll E_{gap}$, therefore,
both the background and the dynamical modes are in the quantum regime. We expect the quantum
corrections to modify the dynamics, break the scale invariance and most probably quench the
influence of the AdS$_2$ poles.
As in Region~3, it is not clear if the low-energy modes have an effective semiclassical
description in this regime, and it will depend crucially on how big the quantum fluctuations are
compared to average values.

We find
\be \label{eq:smallTintro}
\text{Im} \, \CG^{\Delta=1}_{_R} (\omega)
\=  \dfrac{\sqrt{\omega}}{\sqrt{2\pi^2 C}}  + \dots  \,\;\;,\;\;\;  C\equiv {1\over E_{gap}}\ .
\ee
Here the quantum corrections changing the dimension of the IR operator from $\Delta=1$ to $\Delta={1\over 2}$ a relevant value in one dimension.
The details, including the nature of the corrections to the above formula are given in
Equation~\eqref{eq:Tlomquantb} and the surrounding discussion.

\medskip

The result for the imaginary part of the retarded IR Green's function for the regions 4--6
can be summarized in the single formula~\eqref{eq:Tlomquant}, which can be approximated
by~\eqref{eq:Tlomquanta} and~\eqref{eq:Tlomquantb} in the respective regimes mentioned above.

\bigskip

The regions~3 and~6 are the most interesting ones.
The results~\eqref{eq:smallomintro},~\eqref{eq:smallTintro} show
strong deviations from the semiclassical formula. In particular, the semiclassical result~$\omega$ is enhanced
by~$1/\sqrt{C\omega}$ when~$T \ll \omega$ and by~$1/\sqrt{CT}$ when~$\omega \ll CT^{2}$.

\end{enumerate}

\vspace{-0.2cm}

In Section \ref{sec:queta} we focus on the shear viscosity $\eta$, defined via the imaginary part of the retarded Green's
function  in the limit $\omega\rightarrow 0$, as well as its ratio with the entropy density~$\eta/s$.
In the semi-classical regime,  $CT\gg1$, the ratio $\eta^{qu}/\eta^{s.c.}$
behaves as
\footnote{Quantum corrections to the shear viscosity have been calculated in ~\cite{Ge:2018lzo} in the context of the study of coupled SYK models as models of non-Fermi liquids.}
$${\eta^{qu}\over \eta^{s.c.}}\sim1+\tO\left({1\over CT}\right),$$
whereas in the quantum regime, $CT \ll 1$, it behaves
as
$${\eta^{qu}\over \eta^{s.c.}}\sim \sqrt{2/\pi^3 CT} +\tO(\sqrt{CT}).$$
Systematic expansions to the formula in different regimes can and have been calculated.
The details are given in Equations~\eqref{eq:ImGres},~\eqref{eq:ImGreslims},  and in Figure~\ref{fig:ImGres}.
As a check of our results, we note that the quantum corrected~$\eta$,
is in agreement with the quantum corrected absorption cross section computed for fixed temperature in~\cite{Biggs:2025nzs}.

The quantum-corrected~$\eta/s$ behaves as follows: for large values of $CT$, $\eta/s$ asymptotes, as expected,
to the semi-classical universal value $1/4\pi$, whereas as we lower the values of $CT$, there is a dip towards a
minimum value and then it diverges
in the limit $CT\rightarrow 0$. The details are given in
Equation~\eqref{eq:EtaovsAns} and Figure~\ref{fig:EtaovsPlot}.

\medskip

Finally, we point out the similarity between the behaviour of the quantum corrected shear viscosity~$\eta^{qu}$ as
the temperature approaches zero and the behaviour of classical glassy systems in the same limit.
In the latter case, the analogue of the Schwarzian modes corresponds to the correlated motion of larger and
larger parts of the system leading to an increased sensitivity to shear deformations.
The unlimited rise of $\eta/s$ as the temperature asymptotes to zero, in particular,
suggests that the low-energy dynamics becomes glassy.

\medskip

We close the main part of our paper with an outlook in Section~\ref{outlook}, in which we discuss various interesting
questions and potential extensions of the present work. This is followed by two appendices, in which we give various
details regarding the computations of the main results. In particular, in Appendix~\ref{app:waveeqn} we discuss the
fluctuation equation for a massless neutral scalar field in the near-extremal limit of the asymptotically $AdS_{4}$
Reisner-Nordstr\"{o}m  black brane. In appendix \ref{B}, we discuss arguments that indicate that the IR dimension of the shear part of the energy-momentum tensor has IR dimension $\Delta=1$.
In Appendix~\ref{app:intcalcs} we present details of the calculations underlying
the evaluation of the Green's function in Section~\ref{sec:quGreen}, in the various regimes of approximation.

\bigskip

\ndt {\bf Note added: } While the present paper was in preparation, the
papers~\cite{PandoZayas:2025snm,Nian:2025oei,Cremonini:2025yqe} appeared on the arxiv,
which have some overlap in the questions addressed with the present paper.
We found it difficult to compare our results with~\cite{PandoZayas:2025snm,Nian:2025oei}.
The paper~\cite{Gouteraux:2025exs} also appeared a day before the present paper appeared on the arxiv.
Our results differ from these results because in the present paper we study an operator with~$\Delta=1$
in the near-AdS$_2$ region, while~\cite{Gouteraux:2025exs} studies an operator with~$\Delta=0$.
We thank the authors of~\cite{Gouteraux:2025exs} for detailed discussions.

\section{Near-extremal black branes in AdS$_4$ \label{sec:scbranes}}

In this section, we review the black brane solution in asymptotically~AdS$_4$ space.
We consider the near-extremal limit in the semi-classical approximation, and observe the
appearance of the energy scale~$E_\text{gap}$ that governs the fluctuations of the nearly
gapless modes of the theory.

\vskip 0.2cm

\ndt {\bf The black brane solution}

\vskip 0.2cm

\ndt We consider the Einstein-Maxwell theory in AdS$_4$ space,
as defined by the following action,
\be \label{eq:EMact}
S_\text{E-M} \= \frac{1}{16 \pi G} \int \, \dd^4 x \, \sqrt{-g} \, \Bigl( R - F^2 + \frac{6}{L^2} \Bigr) \,.
\ee
We follow the conventions of~\cite{Chamblin:1999tk}.

The metric of the charged black brane solution of this theory is given by
\be \label{eq:BBmetric}
ds^2 \= -f(r) \,\dd t^2 + \frac{\dd r^2}{f(r)} + \frac{r^2}{L^2} \, 4 \pi V_2 \,  \bigl(\dd x_1^2 + \dd x_2^2 \bigr) \,,
\ee
with the blackening factor
\be  \label{eq:deff}
f(r) \= \frac{r^2}{L^2} - \frac{2 G M}{r} + \frac{G Q^2}{r^2} \,.
\ee
The gauge field of the solution is given by
\be
A \=  \Bigl(\, \mu L -\frac{\sqrt{G} \,Q}{r} \, \Bigr) \dd t\,.
\ee

\medskip

As~$r \to \infty$, the metric asymptotically locally approaches that of~AdS$_4$.
At conformal infinity one has three-dimensional flat space labelled by~$(t,x_1,x_2)$.
Here~$x_1,x_2$ are dimensionless coordinates of periodicity~1 and,
as displayed in the metric~\eqref{eq:BBmetric},
the space covered by~$(x_1,x_2)$ is a torus of area~$4 \pi V_2$.

\medskip

The total energy, and the energy density are
\be
E_\text{tot}\= \frac{M \, V_2}{L^2}  \,, \qquad
\rho \= \frac{M}{4 \pi L^2} \,.
\ee
The total charge and the charge density are given by
\be
Q_\text{tot}  \=  \frac{Q \, V_2}{L^2} \,, \qquad \rho_{Q} \= \frac{Q \, V_2}{4 \pi L^2}\,.
\ee

\bigskip

The polynomial~$f(r)$ has two real positive roots, i.e.~$f(r_\pm)=0$, $r_+\ge r_-$, and~$r_+$ and~$r_-$
are the locations of the outer and inner horizons, respectively.
The temperature and the chemical potential, as measured from asymptotic infinity,
can be derived from demanding smoothness in the Euclidean theory of the metric and the gauge field at the outer horizon,
which leads to the following expressions,
\be
T \= \frac{1}{4 \pi} \, f'(r_+) \,, \qquad \mu \= \frac{\sqrt{G}}{L}{Q\over r_+}\,.
\ee

In formulating the AdS$_4$/CFT$_3$ correspondence, we fix the chemical potential~$\muext$
and the temperature~$T$. The other parameters of the solutions can be determined
in terms of~$(\muext, T)$ by the above expressions.

\vskip 0.4cm

\ndt{\bf The near-extremal limit}

\vskip 0.1cm

\ndt The near-extremal condition is~$T \ll \muext$.
In the extremal limit, the locations of the two horizons coincide, i.e.~$r_+=r_-\equiv \rext$,
and the temperature vanishes. In the near-extremal solution, the parameters can be
expressed as the following expansions,

\be \label{eq:rMQT}
\begin{split}
&r_+ (\muext,T)  \=   \rext   \Bigl( 1+ \frac{2 \pi}{\sqrt{3}}  \frac{T}{\mu } + \frac{2 \pi^2}{3} \frac{T^2}{\mu^2} + \dots  \Bigr) \,, \quad
r_- (\muext,T)  \=   \rext   \Bigl( 1 + 0 - \frac{2 \pi^2}{9} \frac{T^2}{\mu^2} + \dots  \Bigr) \,, \\
& Q (\muext,T)  \=   \Qext   \Bigl( 1+ \frac{2 \pi}{\sqrt{3}}  \frac{T}{\mu } + \frac{2 \pi^2}{3} \frac{T^2}{\mu^2} + \dots  \Bigr)\,, \quad
M (\muext,T) \=  \Mext  \Bigl( 1+ \sqrt{3} \, \pi \frac{T}{\mu } +2 \pi^2 \frac{T^2}{\mu^2}+ \dots \Bigr) \,,
\end{split}
\ee
with the parameters of the extremal solution given by
\be \label{eq:extrrQM}
   \Mext \= \frac{2\,\muext^3\,L^4}{3 \sqrt{3}\, G} \,, \qquad
      \Qext \= \frac{\muext^2 \, L^3}{\sqrt{3 \, G}} \,, \qquad
   \rext \= \frac{\muext \, L^2}{\sqrt3} \,.
\ee
The blackening factor in the extremal limit takes the following form
\be  \label{eq:fext}
f^\text{ext}(r)  \= \frac{(r-\rext)^2 \, \bigl(r^2+2 \rext \,r+ 3 \rext^2 \bigr)}{L^2 \, r^2} \,.
\ee

\vskip 0.4cm

\ndt {\bf Semi-classical near-extremal thermodynamics of the black brane}

\vskip 0.1cm

\ndt The semiclassical Bekenstein-Hawking entropy of the near-extremal black brane is given by
\be \label{eq:STsc}
S^\text{s.c.}  (\muext,T)
\= \frac{\pi \, V_2 \, r_+^2}{L^2 \, G}
\= \biggl( \, \frac{\pi}{3} \, \frac{ V_2 \, \muext^2  \, L^2}{G} +
4\pi^2 \frac{T}{E_\text{gap}} \, \biggr) \Bigl( 1+ \tO\bigl(T/\mu \bigr) \Bigr)\,,
\ee
with
\be  \label{eq:defEgap}
E_\text{gap}
\= \frac{3 \sqrt{3} \, G}{\muext \, V_2 \, L^2}
\= \frac{3 \sqrt{3}}{\muext \, V_2 \, \CN}
\;.
\ee
Here we have defined the parameter
\be
{\cal N} \; \equiv \; {L^2\over G} \,,
\label{nu}\ee
which scales as a positive power of
the number of degrees of freedom of the dual QFT.
The proportionality factor depends on the embedding of our simple Einstein-Maxwell sector to a full supergravity.

The semiclassical entropy density at extremality is given by
\be \label{eq:defsBH}
s^\text{s.c.}
\= \frac{S^\text{s.c.} (\muext,0)}{4\pi V_2}
\= \frac{\rext^2}{4 \, L^2 \, G}
\= \frac{\muext^2\,\CN}{12}
\,.
\ee

\medskip

Note that the thermodynamics of the black brane differ in detail from those of the four-dimensional black hole.
In particular, compared to the spherically symmetric black hole, there is a new parameter~$V_2$
that affects the expression for the volume of the black brane. As we observe below, in the limit of infinite~$V_2$,
quantum effects are suppressed at any non-zero energy, and henceforth we keep~$V_2$ finite.

\vskip 0.4cm

\ndt{\bf The near-extremal near-horizon limit and the origin of the Schwarzian mode}

\vskip 0.1cm

\ndt The near-extremal limit can be encoded in a small dimensionless parameter defined as
\be
\ve \; \equiv \; \frac{r_+-r_-}{\rext} \; \ll \; 1 \,,
\ee
which can be traded for the physical temperature~$T$
 at the asymptotic AdS$_4$ region using the relations
\be \label{eq:Tverel}
T \= \frac{\sqrt{3}}{2  \pi}  \,  \muext\, \ve  + \tO(\ve^2) \quad \Longleftrightarrow \quad
\ve \= \frac{2 \pi}{\sqrt{3}} \, \frac{T}{\muext} + \tO((T/\muext)^2) \,.
\ee
The near-extremal limit can be expressed as~$\ve \ll 1$ or, equivalently, as $\frac{T}{\mu} \ll 1$,
and the extremal brane solution has~$\ve = 0$ or, equivalently,~$T=0$.

\bigskip

It is useful to define the dimensionless coordinates~$(\tau,\rho)$,
\be \label{eq:deftaurho}
\tau \= \frac{6 r_\star}{L^2} \, \ve \, t\,, \qquad
\rho \= \frac{1}{\ve} \frac{r-r_+}{\rext}\,,
\ee
in terms of which the black brane solution \eqref{eq:BBmetric} admits the expansion
\be \label{eq:Tcorrmetric}
\begin{split}
\dd s^2 &\=  \frac{L^2}{6} \Bigl( -\rho(1+\rho) \dd \tau^2+\frac{\dd \rho^2}{\rho(1+\rho)}  \Bigr)
+ {\rext^2\over L^2} \, 4\pi V_2 \bigl( \dd x_1^2 + \dd x_2^2 \bigr) +  \\
& \qquad + \biggl( \, \frac{2\pi}{9\sqrt{3}}\frac{T}{\mu} \,  L^2 \, \Bigl( \rho(1+3\rho+2\rho^2)  \, \dd \tau^2
+ \frac{1+2\rho}{\rho(1+\rho)} \, \dd \rho^2 \Bigr) +\\
& \qquad \qquad  \quad  + 16 \pi^2\frac{T}{E_\text{gap}} \, (1+\rho) \left(\dd x_1^2+\dd x_2^2\right) \biggr)
\Bigl(1+ \tO(T/\mu) \Bigr) \,,
\end{split}
\ee
with~$E_\text{gap}$ given in~\eqref{eq:defEgap}.

The first line in~\eqref{eq:Tcorrmetric} is the near-horizon  AdS$_2 \times \, T^2$ solution
in Rindler coordinates.
Upon Euclidean rotation of the geometry, we obtain a hyperbolic disk.
Note that we have taken the extremal limit in a manner as in~\cite{Sen:2008yk,Maldacena:1998uz,Faulkner:2009wj},
so that the outer horizon of the black hole ($\rho=0$) is at finite radial distance from any point in the interior of AdS$_2$.
Hence we refer to this as the AdS$_2$ BH.
It is, however, important to note that the temperature of this geometry
as measured from the asymptotic AdS$_4$ boundary is zero.
Moreover, backreacting perturbations inside this spacetime do not respect the AdS$_2$
boundary~\cite{Maldacena:1998uz}. On the other hand, including the second and third line
produces a \emph{nearly}-AdS$_2$ spacetime which allows for consistent
backreaction~\cite{Almheiri:2014cka, Maldacena:2016upp, Hadar:2020kry}.

The second and third lines in~\eqref{eq:Tcorrmetric} contain the first small-temperature
corrections to the AdS$_2$ geometry.
Note that while the second line is suppressed by~$T/\mu$,
the third line is only suppressed by~$T/E_\text{gap}$.
The metric mode contained in the third line (which we call $\delta g$)
controls the size of the transverse space, and grows linearly
in the coordinate~$\rho$, i.e.~exponentially in the proper distance towards asymptotic infinity.
{The coefficient of}
this exponentially growing mode~$\delta g$ is identified as
the coupling of the one-dimensional Schwarzian mode living on the
boundary of Euclidean~AdS$_2$~\cite{Almheiri:2016fws,Maldacena:2016upp}.

We now briefly recall the argument underlying the appearance of the Schwarzian in the four-dimensional
theory~\cite{Nayak:2018qej, Moitra:2018jqs, Castro:2018ffi,Sachdev:2019bjn, Iliesiu:2020qvm, Heydeman:2020hhw,
Iliesiu:2022onk, Iliesiu:2022kny}.
We follow the treatment of~\cite{Iliesiu:2022onk}, making modifications due to the fact that~\cite{Iliesiu:2022onk}
studied AdS$_2 \times S^2$ in a theory with no cosmological constant, while in the present context we
consider~AdS$_2 \times T^2$ in a theory with cosmological constant.
When we expand the metric into off-shell fluctuations~$h$, the
fact that~$\delta g$ in~\eqref{eq:Tcorrmetric} solves the equation of motion implies that there are no couplings of the
form~$\delta g \, h$ and the first non-trivial couplings are of the form~$\delta g \, h^2$. The values of the Lagrangian
of generic perturbations~$h$ at~$T=0$ (i.e.~the KK masses of the perturbations) have the scale of the curvature,
which is~$1/L^2$ for generic AdS$_2$ fluctuations, and~$L^2/\rext^2 V_2$ for the fluctuations of the transverse space.
These KK masses obtain additive corrections with coefficients proportional to~$T$ at small temperature.
Integrating out these modes leads to corresponding corrections proportional to~$T$
to their effective action. In particular, since their extremal action is non-zero, the~\hbox{$T=0$} limit is smooth.

The exception comes from special off-shell fluctuations of the metric which have zero Laplacian on~AdS$_2$ \cite{Camporesi:1995fb}.
At small~$T$, these modes gain an action proportional to~$T/E_\text{gap}$.
The~$T \to 0$ limit of these light modes is not smooth in the effective theory, and they can give rise to strong quantum effects.
These are precisely the Fourier components of the Schwarzian field\footnote{We expect that the Einstein-Maxwell-$\Lambda$
action for this off-shell mode gives  the Schwarzian action, similar to analogous calculations in flat space~\cite{Iliesiu:2022onk}.
It would be nice to explicitly verify this. }.
We discuss the results for the partition function and the two-point function
of the Schwarzian theory as applied to our problem in Section~\ref{sec:quGreen}.
Before that, in the following section, we turn to
the semi-classical calculations of the holographic Green's functions.

\section{Semiclassical Green's functions \label{sec:SCGreen} }

In this section we briefly recall the semiclassical holographic result for transport coefficients,
focussing on the shear viscosity~$\eta$.
The starting point is the holographic relation~\cite{Policastro:2001yc,Policastro:2002se}
between the retarded Green's functions of any operator in the boundary theory,
and the solution of the wave equation for the field in the bulk theory that couples to the given operator.

In the bulk dual discussed in Section~\ref{sec:scbranes},
the retarded UV Green's function corresponds to the
asymptotic~AdS$_4$ region, which we denote by~$G_{R}(\omega,\vec{q})$.
The corresponding IR Green's functions are associated to the~AdS$_2$ region deep down the throat,
which we denote by~$\CG^{\vec{q}}_{_R}(\omega)$.

The smooth solution of the wave equation in the bulk theory relates the UV Green's function
to the IR Green's functions.

In this paper we focus on the transport coefficient, which corresponds to the modes
with~$\vec{q}=0$ and small frequency $\omega$.
We denote the corresponding IR Green's function simply by~$\CG_{_R}(\omega)$.

We review the solution to the wave equation for a massless neutral scalar field at very low temperatures
in Appendix~\ref{app:waveeqn}. In the semiclassical approximation, we can read off the value of
the corresponding UV Green's function from the asymptotic behavior of the wave,
following the prescription of~\cite{Son:2002sd}.
In particular, it is controlled by the ratio of the two branches of solutions corresponding to the source
and vev, respectively.

\vskip 0.2cm

\ndt{\bf Effective scalar equations from Einstein-Maxwell theory}

\vskip 0.1cm

\ndt We are interested in the two-point correlator of the transverse stress tensor mode~$T_{xy}$
in the boundary theory. In particular, we consider the retarded Green's function, which we
denote by $G_{R\,xy,xy}(\omega,\vec{q})$.
In the dual bulk AdS$_4$ theory, this is related to two-point function of the graviton~$h_{xy}$ with itself.
In the full gravitational theory, the equation for the metric fluctuation mixes with the equation governing the gauge field,
which is, in general, complicated. However, as shown in~\cite{Edalati:2009bi,Edalati:2010hk},
upon studying the equation in the corresponding gauge-invariant variable using the Kodama-Ishibashi master-field
formalism~\cite{Kodama:2003kk}, one finds a simplification in that
one obtains precisely the equation of a massless, neutral scalar field in the bulk AdS$_4$.

As is well-known, low-energy fields in the bulk AdS$_4$ couple to (or source) operators in the
AdS$_2$ theory in the deep IR region. The~$\vec{q}=0$ mode of the massless, neutral scalar field
that we consider here couples to an operator with~$\Delta=1$ in the AdS$_2$ theory.
In this context, the holographic Green's function calculation can be described as follows.
We begin with two insertions at the asymptotic~AdS$_4$ boundary, which source two bulk fields.
These two fields propagate to the IR and, upon hitting the~AdS$_2$ boundary, source two operators in~AdS$_2$.
The correlator of these two operators in AdS$_2$ is then computed in the semi-classical theory by
studying the propagation in AdS$_2$.

\vskip 0.2cm

\ndt{\bf IR limit and shear viscosity}

\vskip 0.1cm

\ndt Carrying out the steps mentioned above, following Appendix~\ref{app:waveeqn}, one obtains the relation
between the UV and IR Green's functions~\cite{Faulkner:2009wj,Edalati:2009bi,Edalati:2010hk,Davison:2013bxa}:
\be \label{eq:GUVIRrel}
G^{\text{s.c.}}_{R \, xy,xy}(\omega,\vec{q}=0)
\= -\frac{\rext^2}{16 \pi G L^2} \frac{\CG^{\text{s.c.}}_{_R}(\omega)}{1+ \xi \, L^2 \,
\CG^{\text{s.c.}}_{_R}(\omega)}
\= -\frac{\mu^2 \, \CN}{48 \pi} \, \frac{\CG^{\text{s.c.}}_{_R}(\omega)}{1+ \xi \, L^2 \,
\CG^{\text{s.c.}}_{_R}(\omega)}  \,,
\ee
with $\xi=\frac{-\frac{\pi }{2}+\sqrt{2} \log 6+\tan ^{-1}\sqrt{2}}{18 \sqrt{2} \, \rext}$
and
\be\label{eq:AdS2Green}
 \CG^{\text{s.c.}}_{_R}(\omega) \=  i \omega  \,.
\ee
To leading order in~$\omega$ small, we have
\be \label{eq:GUVIRlead}
G^{\text{s.c.}}_{R\,xy,xy}(\omega,\vec{q}=0)
\= -\frac{\mu^2 \, \CN}{48 \pi} \, \CG^{\text{s.c.}}_{_R}(\omega) \bigl(1+ \tO(\omega/\mu) \bigr) \,.
\ee

\ndt  To study the full hydrodynamic-like expansion at fixed~$\mu$ we need to calculate the connection
coefficients of the wave equation to a higher order in the small~$\omega$ expansion.
However, the above results at first order already lead to an expression for the shear viscosity
using Kubo's formula. In the semi-classical theory we have
\be \label{eq:etarrel}
\eta^\text{s.c.} \= - \lim_{\omega \to 0} \, \frac{1}{\omega}\, \text{Im} \, G^{\text{s.c.}}_{R \,xy,xy}(\omega,\vec{q}=0)
\= \frac{\muext^2 \, \CN}{48 \pi}  \, \lim_{\omega \to 0} \, \frac{1}{\omega}\, \text{Im} \,  \CG^{\text{s.c.}}_R(\omega)
\= \frac{\muext^2 \, \CN}{48 \pi} \,.
\ee
This leads to the famous relation for the shear viscosity in terms of the semi-classical entropy density~\eqref{eq:defsBH},
\be \label{eq:etasc}
\eta^\text{s.c.}  \= \frac{1}{4\pi} s^\text{s.c.}  \= \frac{\muext^2 \, \CN}{48 \pi}   \,,
\ee
where, in the second equality, we have expressed the semi-classical shear viscosity in terms of
the fixed chemical potential using the relation~\eqref{eq:extrrQM}.

\bigskip

The above equations are derived in the extreme low-temperature limit, in which we obtain an~AdS$_2$
IR region.
As mentioned in the introduction, the AdS$_2$ region contains a set of gapless modes which have
strong quantum fluctuations.
In the following sections we re-analyze the above Green's functions in the quantum theory.
In order to do so, we turn on a small temperature.
This affects the above results in multiple ways.
Firstly, even within the semiclassical approximation, it changes the extremal radius~$\rext$ to
the small-temperature radius~$r_+$
in the above results~\eqref{eq:GUVIRlead}, \eqref{eq:etarrel}, as explained in~\cite{Iqbal:2008by}.
Secondly, the temperature regulates the gapless fluctuations in the IR, so that we can perform the path integral
in the quantum theory in the near-AdS$_2$ region.
Within the quantum theory, there are two effects:
\begin{enumerate}
\item[(a)] the entropy density is modified compared to the semiclassical
result and, in fact, the quantum entropy diminishes as~$T \to 0$~\cite{Iliesiu:2020qvm}, and
\item[(b)] the IR Green's function is also modified, and we need to calculate the quantum Green's function.
\end{enumerate}
This is what we turn to in the following section.

\section{Quantum Green's functions \label{sec:quGreen}}

In this section, we study the quantum corrections to the IR Green's function in the nearly AdS$_2$
region that appears deep inside the throat at small temperatures. Throughout this
section and the following one, we analyze the regime~$\omega, T \ll \mu$.
As we discussed in the introduction, there is
another scale~$E_\text{gap}$ that controls the quantum fluctuations.

The IR theory can be described in terms of the nearly-AdS$_2$/nearly-CFT$_1$ correspondence.
In this set-up, consider a real operator~$\mathcal{O}_\Delta(t)$
of conformal dimension~$\Delta$ inserted at two points~$t=t_1$, $t=t_2$ at the boundary.
Using time-translation invariance we can set one of the points to be at~$t=0$.
The real-time Wightman two-point function of this operator is written
as~$\langle  \mathcal{O}_\Delta(t) \, \mathcal{O}_\Delta(0) \rangle$
and the frequency-domain Wightman function is given by its Fourier transform\footnote{The normalization
of the Fourier transform is such that the Green's functions in the time-domain and the frequency-domain,
quoted below from different sources in the literature, are consistent with each other.},
\be
\CG^\Delta (\omega) \=
 \frac{1}{\pi} \,\int_{-\infty}^\infty \, dt \, e^{i \omega t} \,
\langle  \mathcal{O}_\Delta(t) \, \mathcal{O}_\Delta(0) \rangle \,.
\ee
Our focus here is on the imaginary part\footnote{Note that the full retarded Green's function can be
reconstructed from the knowledge of the imaginary part, using the Kramers-Kronig relation, but
we shall not study it here.} of the retarded Green's function at temperature~$T=1/\beta$, which can be expressed
in terms of the Wightman function, $\CG^\Delta (\omega)$, as
\be \label{eq:FDthm}
\text{Im} \, \CG^\Delta_{_R} (\omega) \= \frac12 \bigl( 1-e^{-\beta \omega} \bigr) \, \CG^\Delta (\omega) \,.
\ee
The relation~\eqref{eq:FDthm}, which is the fluctuation-dissipation theorem for the thermal enseble, can be explicitly checked by separately
calculating the two sides in the quantum Schwarzian theory.

\vskip 0.4cm

\subsection {The quantum Schwarzian theory}

\vskip 0.1cm

\ndt As explained at the end of Section~\ref{sec:scbranes}, we replace the semi-classical Wightman function
in the AdS$_2$ theory by the quantum result obtained by including the effect of the quantum fluctuations
of the Schwarzian mode living at the boundary of~AdS$_2$~\cite{Mertens:2017mtv,Lam:2018pvp,Yang:2018gdb}.
We refer the reader to the review~\cite{Mertens:2022irh} for an exposition and detailed references,
and briefly summarize the results here.

To begin with, it is important to note the scales governing the quantum fluctuations.
As noted in Section~\ref{sec:scbranes}, in the semi-classical approximation the first corrections to the
zero-temperature results appear with strength~$T/\mu$. The quantum corrections, on the other hand,
are governed by the (inverse) coupling~$CT$  of the Schwarzian theory, where
\be
C \= \frac{1}{E_\text{gap}} \= \frac{\muext \, V_2 \, \CN}{3 \sqrt{3}} \,.
\label{c}\ee
The relevant energy scale where the Schwarzian mode becomes strongly coupled therefore scales as~$1/\mu$
(keeping all other scales fixed). This is much smaller than~$\mu$ as~$\mu$ becomes large.

When~$E_\text{gap} \ll T \ll \mu$, one can use the saddle-point approximation to the Schwarzian path integral.
As we increase~$T$, these results match the semi-classical approximation to the full black hole.
On the other hand, when~$T \lesssim E_\text{gap}$, we cannot rely on the saddle-point approximation and
need to perform the exact path integral. This result can be trusted up to an exponentially small energy scale~$\exp(-S_0)$,
where~$S_0$ is the semi-classical extremal entropy of the black hole,
at which point non-perturbative effects can kick in. (This is the scale at which
the genuine discreteness of the quantum theory should be seen.)
The bottom line is that the Schwarzian theory approximates the physics well in a range of energies~$e^{-S_0} \ll T \ll \mu$.

\bigskip

\vskip 0.3cm

\ndt \textit{The entropy and the entropy density}

\vskip 0.1cm

\ndt The result for the quantum partition function of the Schwarzian theory in
 the fixed charge ensemble (see e.g.~\cite{Mertens:2022irh}) is given by
\be \label{eq:Zresult}
Z(T) \= \frac{\bigl(C T  \bigr)^{\frac32} }{\sqrt{2 \pi}}
\exp\Bigl(S_0  + 2 \pi^2 CT \Bigr)   \,.
\ee
where
$$S_0\equiv S^\text{s.c.}  (\muext,0)=\frac{\pi}{3} \,  V_2 \, \muext^2 \, \CN $$
is the semi classical extremal entropy.
Note that the partition function of the Schwarzian theory quoted  above, is
the partition function of the quantum states above extremality for a black hole of fixed charge.
\footnote{In particular, the extremal entropy term~$S_0$ is shown in the expression~\eqref{eq:Zresult}, but
there are other terms proportional to~$\beta$
which are present in the canonical ensemble that are suppressed in the formula.
These terms correspond to the zero-point energy and charge of the black hole, and it is consistent to
carry these terms through all the following expressions, which we do implicitly without writing them.}

\begin{figure}[h!]
        \centering
        \includegraphics[scale=1.2]{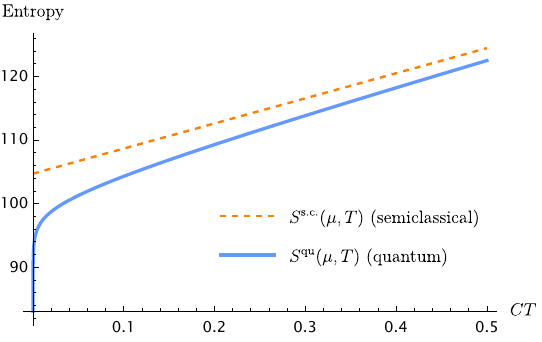}
    \caption{\textit{Semiclassical entropy $S^\text{s.c.}$  and quantum entropy $S^\text{qu}$ as a function of~$CT$.
    The quantum entropy is calculated in the exact Schwarzian theory as in~\cite{Iliesiu:2020qvm}}.
    In this plot we choose~$\mu=10$.
    At large values of~$CT$, the quantum entropy reaches the semiclassical limit, while it approaches zero at small~$CT$.
    It can be trusted at the lower end when~$CT$ is much larger than the cutoff scale when the above quantum entropy curve hits zero.
    This scale ($\approx e^{-\frac23 S_0} \approx 10^{-29}$) is where non-perturbative effects come into play.
   }
    \label{fig:SquPlot}
\end{figure}

Further, in the black brane set-up that we study, we use the grand-canonical ensemble at the asymptotic~AdS$_4$ region.
This means that the exponent of~\eqref{eq:Zresult} is not quite the same as the semiclassical low-temperature entropy
formula~\eqref{eq:STsc} which we repeat here for convenience,
\be \label{eq:STsc1}
S^\text{s.c.}  (\muext,T)
\= \biggl( \, \frac{\pi}{3} \, V_2 \, \muext^2  \, \CN +
4\pi^2 \frac{T}{E_\text{gap}} \, \biggr) \Bigl( 1+ \tO\bigl(T/\mu \bigr) \Bigr)\,,
\ee
indeed, the Gibbs free energy in the grand-canonical ensemble in the semi-classical approximation
with~$M(\muext,T)$, $Q(\muext,T)$, and~$S^\text{s.c.}(\muext,T)$ given in~\eqref{eq:rMQT}, \eqref{eq:STsc1} is
\be \label{eq:GmuT}
\begin{split}
-\beta \, G(\muext,T) & \= -\beta M_\text{tot}(\muext,T) + \beta \muext  L \, Q_\text{tot}(\muext,T) + S^\text{s.c.}(\muext,T) \\
& \=\bigl( -\beta \Mext + \beta \mu  L \,\Qext  \bigr) \frac{V_2}{L^2}+ S^\text{s.c.}(\muext,0)  + 2 \pi^2 CT \,.
\end{split}
\ee
We observe that, after dropping the first two terms on the right-hand side,  the above expression
agrees precisely with the right-hand side of~\eqref{eq:Zresult}.
The implication (that we use in our calculation of~$\eta/s$ below) is that we should use
the following expression for the quantum entropy~\cite{Iliesiu:2020qvm},
\be
\begin{split}
S^\text{qu}(\mu,T) & \=  S^\text{s.c.}(\muext,T) +  \log \Bigl( \frac{(C T)^{\frac32} }{\sqrt{2 \pi}}\Bigr) \\
& \=  S^\text{s.c.}(\muext,0) +4 \pi^2 CT + \frac32 \log (CT) -\frac12 \log (2 \pi) \,.
\end{split}
\ee
We plot this quantum entropy as a function of temperature in Figure~\ref{fig:SquPlot}.

\medskip

Using the above formulae, we also express the entropy density in the following form,
\be \label{eq:squ}
\frac{s^\text{qu}(\mu,T)}{s^\text{s.c.}(\muext,T)} \=
\frac{S^\text{qu}(\mu,T)}{S^\text{s.c.}(\muext,T)} \=
1 + \frac{\log \bigl((C T)^{\frac32} /\sqrt{2 \pi} \,\bigr)}{S^\text{s.c.}(\muext,T)} \,,
\ee
which is useful in the discussion below.

\medskip

There are two technical points to note. Firstly, when~
$$CT \approx \exp(-\frac23 S^\text{s.c.}(\muext,0)),$$
 the
quantum entropy vanishes. This is the scale mentioned above, which provides a lower cutoff to the
regime of validity of the Schwarzian description of the theory.
Secondly, we have to make a choice of ensemble in the Schwarzian theory. In our problem, it is clear that we should keep
temperature (and not energy) fixed in the nearly AdS$_2$ theory, but there is also the choice of chemical potential vs charge.
Depending on this choice, the~$U(1)$ Schwarzian-like mode~\cite{Sachdev:2019bjn} fluctuates or is frozen in the path integral.
However, the fact that we are interested in correlators of an uncharged scalar field means that the~$U(1)$ mode decouples
from our calculations, and will not appear in the normalized partition function that we consider below.

\vskip 0.3cm

\subsection{The two-point function}

The exact two-point Wightman function in the Schwarzian theory
is given by the following expression~\cite{Mertens:2022irh},
\footnote{The expressions in~\cite{Mertens:2022irh} include a regulator of the
form~$\exp \bigl(- \varepsilon \frac{k_1^2}{2 C} - \varepsilon \frac{k_2^2}{2 C} \bigr)$ with  $\varepsilon \searrow 0$
to ensure convergence of the integral, but we do not seem to need it below.}
\be \label{eq:OOcor}
\begin{split}
\langle  \mathcal{O}_\Delta(t) \, \mathcal{O}_\Delta(0) \rangle & \=
\frac{e^{S_0}}{Z(T) \, 8 \pi^4 \, (2C)^{2\Delta} \, \Gamma(2 \Delta)} \, \times  \\
&  \qquad \int_0^\infty \prod_{i=1,2} d k_i^2 \, \sinh (2 \pi k_i) \,
e^{- i t \frac{k_1^2}{2 C} - (\beta - it) \frac{k_2^2}{2 C}} \,
\prod\limits_{\sigma_1,\sigma_2=\pm 1}
\Gamma(\Delta+ i \sigma_1 k_1+ i \sigma_2 k_2) \,.
\end{split}
\ee
Note that we have normalized the 2-point function as given in~\cite{Mertens:2022irh} by the partition
function~$Z(T)$ as appropriate for the study of holographic correlators.

\medskip

\ndt We are eventually interested in~$\text{Im} \, \CG_{_R}(\omega)$.
Given that the operator~$\mathcal{O}$ is real, this is an odd function of $\omega$,
and hence it is enough to consider~$\omega >0$. 	
Moving to frequency-domain, using the Fourier transform
\be
\begin{split}
\frac{1}{\pi} \, \int_{-\infty}^\infty dt \, e^{i\omega t}\, e^{-i t (k_1^2 - k_2^2)/2C}
\= 2\delta \Bigl(\omega-\frac{k_1^2-k_2^2}{2C} \Bigr)
\= 4 C \delta \bigl(k_1^2- (k_2^2+2C\omega) \bigr)  \,,
\end{split}
\ee
the double integral in the expression~\eqref{eq:OOcor}
collapses to a single integral for the Wightman function in frequency space.
As a result, we obtain the following frequency space Wightman function\footnote{Equation (\ref{eq:Gintk}) is valid for $\omega>0$. When $\omega<0$ we obtain the same expression with $\omega \to |\omega|$ multiplied by a factor of $e^{-\beta|\omega|}$. 
This is the correct expression in order for  $\text{Im} \, \CG_{_R}(\omega)$ in (\ref{eq:FDthm}) to be an odd function of~$\omega$.}:
\be \label{eq:Gintk}
\begin{split}
\CG^\Delta (\omega)  & \=
 \frac{e^{S_0} (2C)}{Z(T) \, 4 \pi^4 \, (2C)^{2\Delta} \, \Gamma(2 \Delta)} \,
 \int_0^\infty d k \, 2k \, \sinh (2 \pi k) \, \sinh \bigl(2 \pi \sqrt{k^2 + 2 C \omega} \bigr) \,
e^{ - \frac{k^2}{2 C T}} \, \times \\
& \qquad \qquad \qquad \qquad  \qquad \qquad  \qquad \qquad
\prod\limits_{\sigma_1,\sigma_2=\pm 1}
\Gamma \bigl(\Delta+ i \sigma_1 \sqrt{k^2 + 2 C \omega}+ i \sigma_2 k \bigr) \,.
\end{split}
\ee

When $\Delta=1$ the products of the Gamma functions in~\eqref{eq:Gintk} simplify, upon using the
property $\Gamma(1-z) \,\Gamma(1+z) = \pi z/\sin(\pi z)$, $z \notin \mathbb{Z}$, as follows,
\be
\prod\limits_{\sigma_1,\sigma_2=\pm 1}
\Gamma \bigl(\Delta+ i \sigma_1 \sqrt{k^2 + 2 C \omega}+ i \sigma_2 k \bigr)
\=  \frac{2 \pi^2 C\omega}{\sinh \bigl(\pi(\sqrt{k^2 +  2C \omega}+k) \bigr)
\sinh \bigl(\pi(\sqrt{k^2 +  2C \omega}-k) \bigr)} \,.
\ee
The two-point function therefore takes the following form
\be \label{eq:Gintk1}
\begin{split}
\CG^{\Delta=1} (\omega)  & \=
 \frac{e^{S_0}\, \omega}{Z(T) \, 2 \pi^2} \,  \int_0^\infty d k \, k \,
e^{ - \frac{k^2}{2 C T}} \, \sinh (2 \pi k) \,\times \\
& \qquad \qquad \qquad \qquad
\frac{\sinh \bigl(2 \pi \sqrt{k^2 + 2 C \omega}\, \bigr)}{\sinh \bigl(\pi(\sqrt{k^2 +  2C \omega}+k) \bigr)
\sinh \bigl(\pi(\sqrt{k^2 +  2C \omega}-k) \bigr)} \,.
\end{split}
\ee

\vskip 0.4cm

\subsection{The Green's function in various regimes\label{regime}}

\vskip 0.1cm

\ndt We briefly analyze the integrand in~\eqref{eq:Gintk1}. As~$k \to 0$, the
integrand vanishes so that the integral converges at the lower end.
As~$k \to \infty$, the damped exponential factor~$e^{-k^2/2CT}$ dominates
the exponential growth of the rest of the integrand (for which the exponent is linear in~$k$),
and so the integral converges at the upper end as well.
In the middle, there is a competition between the damped term,
and the growing terms.
We can check numerically that the integrand always has one peak and an exponential fall off,
and we refer to this shape as a bell, keeping in mind that the bell can be skewed depending on the parameters.

We can first gain some intuition by focussing on the first line of~\eqref{eq:Gintk1}.
When~$CT \gg 1$, there is a peak at large~$k$. In this range, the sinh can be approximated by an exponential, and
the shape is approximately a Gaussian peaked at~$k=2 \pi CT$ with
variance~$CT$.
When~$CT$ is small, there is still a peak of the integrand for small~$k$, but now the peak is no longer Gaussian.
To estimate this, we can approximate $\sinh x \approx x$ to see that the peak is at~$k \approx \sqrt{CT}$.
In fact, as we discuss below, the integral in the first line, as well as related integrals that appear in our analysis,
can be evaluated in terms of simple functions for all values of~$CT$.

As explained in the introduction, the analysis can be split into different regions depending on the
relative strengths of the three relevant scales~$E_\text{gap}$, $\omega$, and~$T$
or, equivalently, $1$, $C\omega$, and~$CT$ (recall that~$C=1/E_\text{gap}$).
The main point of the approximation is that the second line in~\eqref{eq:Gintk1}
can be replaced in the integral by either a constant or a linear function of~$k$ to different orders of accuracy depending
on the regimes of the parameters.
Upon taking this into account, we obtain different analytic approximations to~\eqref{eq:Gintk1}.
We summarize the results below, and present the details in Appendix~\ref{app:intcalcs}.

\medskip

\ndt {\bf Regime I.} \textit{$\omega$ is smaller than~$T$}

\smallskip

\ndt {\bf I a.} \textit{$1, C\omega \ll CT$, i.e.~$T$ is the largest scale and in the semiclassical regime}

\smallskip

In this case, the background is semiclassical.
The arguments of all three $\sinh(\cdot{})$ functions in the second line of the integrand of~\eqref{eq:Gintk1}
can be replaced by~$\frac12 \exp(\cdot{})$.
In making this replacement in the $\sinh$ function in the numerator, as well as the one in
denominator with the relative positive sign in the exponent,
we drop terms with magnitude~$(1-e^{-2\pi k})$, where~$k \approx \pi CT \gg 1$ in the region of the bell.
On the other hand, in the $\sinh$ function in the denominator with the relative negative sign in the exponent,
we have to be more careful since
\be
1-e^{-2\pi (\sqrt{k^2+2C\omega} -k)} \= \frac{2 \pi C \omega}{k} \Bigl( 1+ \tO\Bigl(\frac{C\omega}{k} \Bigr) \Bigr) \,.
\ee
With these approximations, we obtain
\be \label{eq:GomlargeT}
\begin{split}
\CG^{\Delta=1} (\omega)
  \=    2 \, T \, \Bigl( 1 + \frac{1}{4 \pi ^2 C T}\Bigr) +\tO\bigl(\text{max} \bigl( \frac{2C\omega}{CT}  ,  e^{- 2 \pi^2 C T} \, \bigr) \bigr)\,,
 \end{split}
\ee
Within the same approximations, we obtain, for the imaginary part of the retarded Green's function,
\be \label{eq:ImGintklargeT}
\text{Im} \, \CG^{\Delta=1}_{_R} (\omega)
\= \omega \Bigl( 1 + \frac{1}{4 \pi ^2 C T}\Bigr)+\cdots \,.
\ee

\bigskip

\ndt {\bf I b.} \textit{$C\omega \ll 1, CT, (CT)^2$, i.e.~$\omega$ is the smallest scale and is in the quantum regime.}
Note that the {\bf Ib} case and the {\bf Ia} case overlap in the following regime:   $C\omega \ll 1 \ll  CT$.
\smallskip

In this regime we use the fact that~$C\omega$ is the smallest scale to approximate the integral.
Note that we allow~$CT$ to be either small or large compared to~1.

\smallskip

Since~$C\omega$ and~$CT$ can be much smaller than~1, we cannot approximate
the $\sinh$ functions with exponentials and, so, we take a different approach. We
summarize the main points here and discuss the details in Appendix~\ref{app:intcalcs}.
Firstly, we note that the ratio~$ \pi C \omega  f(k,\omega)/k$
is a monotonically decreasing function of~$k$, and approaches~1 asymptotically as~$k \to \infty$.
This indicates that we can approximate~$f$ by the linear function~$\frac{k}{\pi C\omega}$ in the integral.
Then, we show that to the right of the small region~$[0,k_0]$ with~$k_0 = (C\omega)^\frac14$,
the difference $\bigl(f-\frac{k}{\pi C\omega} \bigr)$ can be made arbitrarily small. Further,
in this small region, the effect of replacing~$f$ by~$\frac{k}{\pi C\omega}$ in the integral
can also be made arbitrarily small.
The final result is that,
with an error~$\tO \Bigl(\text{max} \Bigl(\sqrt{C\omega} , \dfrac{(C\omega)^\frac32}{(CT)^2} \Bigr)\Bigr)$, we have
\be \label{eq:Gomsmallom3}
\CG^{\Delta=1} (\omega)   \=    2 \, T  \biggl(
\text{erf}\left(\sqrt{2} \pi  \sqrt{C T} \right) \Bigl( 1 + \frac{1}{4 \pi ^2 C T}\Bigr)
+ \frac{e^{-2 \pi ^2 C T}}{\sqrt{2 \pi^3 \, C T}}  \biggr) \,,
\qquad \qquad  \quad
\ee
\begin{subnumcases}{\; \quad \=}
\quad 2T \Bigl(  1 + \dfrac{1}{4 \pi ^2 CT} + \tO \bigl( e^{-CT} \bigr)  \Bigr) \,, &  \quad $1 \ll CT $\,,  \label{eq:Gomsmallom3a} \\[10pt]
 \dfrac{\sqrt{8 T} }{\sqrt{\pi^3 C}}  \,
\biggl(1 +\dfrac{2 \pi^2 }{3}  \, C T - \dfrac{2 \pi^4}{15} \, (C T)^2 + \dots \biggr)  \,, & \quad  $CT \ll 1$ \,.  \label{eq:Gomsmallom3b}
\end{subnumcases}
With the same approximations, we now take~$ \frac{C\omega}{CT} \ll 1$.
Since, in the regime assumed here~$\sqrt{C\omega} \gg \frac{C\omega}{CT}$, we have, with the same error as above,

\be\label{eq:omlTquant2}
\text{Im} \, \CG^{\Delta=1}_{_R} (\omega)   \=
  \omega  \biggl(  \text{erf}\left(\sqrt{2} \pi  \sqrt{C T} \right) \Bigl( 1 + \frac{1}{4 \pi ^2 C T}\Bigr)
+ \frac{e^{-2 \pi ^2 C T}}{\sqrt{2 \pi^3 \, C T}}  \biggr) \,, \qquad \qquad \qquad \qquad
\ee
\begin{subnumcases}{\; \quad \=}
\quad \omega \Bigl(  1 + \dfrac{1}{4 \pi ^2 CT} + \tO \bigl( e^{-CT} \bigr)  \Bigr) \,,
& $\quad 1 \ll CT \,,$ \label{eq:omlTquant2a}\\[10pt]
 \dfrac{\sqrt{2} \, \omega}{\sqrt{\pi^3 C T}}  \,
\biggl(1 +\dfrac{2 \pi^2 }{3}  \, C T - \dfrac{2 \pi^4}{15} \, (C T)^2 + \dots \biggr)  \,,
&$ \quad CT \ll 1 \,. $ \label{eq:omlTquant2b}
\end{subnumcases}
Note that~\eqref{eq:omlTquant2} summarizes Regime~I completely.
The result of Regime~Ia (which always satisfies~$1 \ll CT$) is~\eqref{eq:omlTquant2a},
and the result of Regime~Ib is~\eqref{eq:omlTquant2a} or~\eqref{eq:omlTquant2b},
depending on whether~$1 \ll CT$ or~$CT \ll 1$. The errors in the various cases are detailed in
the discussion over the last page.

\bigskip

\bigskip

\ndt {\bf Regime II.} \textit{$T$ is smaller than~$\omega$}

\smallskip

In this region, $C\omega$ is much larger than the peak of the integrand in~\eqref{eq:Gintk1}.
Recall that the peak is around~$2 \pi CT$ for~$CT \gg 1$, and around~$\sqrt{CT}$ for~$CT \ll 1$.
So we seek to approximate the function~$f$ appearing in the second line of~\eqref{eq:Gintk1}
in the regime~$k \ll C\omega$.
We have
\be
\begin{split}
\frac{f(k,\omega)}{2\coth \bigl(\pi\sqrt{2C \omega} \bigr) }
& \= \begin{cases}
1 + \tO\bigl( \frac{k}{\sqrt{C\omega}} \bigr) \,,  \qquad  & C\omega \ll 1 \,, \\
1 + \tO\bigl( k \, e^{-\sqrt{C\omega}} \bigr) \,,  \qquad  & C\omega \gg 1 \,.
\end{cases}
\end{split}
\ee
We observe that, as long as~$k^2 \ll C\omega$, the right-hand side is well-approximated by~1.

\bigskip

\ndt {\bf II a.} \textit{$1, CT, (CT)^2 \ll C\omega$}: In this regime the error
is $\tO\bigl( \text{max} \bigl(\sqrt{CT} \, e^{-\sqrt{C\omega}} , CT \, e^{-\sqrt{C\omega}}  \bigr)\bigr)$.

\smallskip

\ndt {\bf II b.} \textit{$ CT \ll C\omega \ll 1$}: In this regime, the error is $\tO\bigl( \sqrt{\frac{CT}{C\omega}} \bigr)$. \\

\medskip

With these errors, the two-point function takes the following form,
\be \label{eq:GomsmallT}
\begin{split}
\CG^{\Delta=1} (\omega)
& \= \frac{e^{S_0}\, \omega \,  \coth \bigl(\pi \sqrt{2 C\omega} \,\bigr)}{Z(T) \,  \pi^2} \,  \int_0^\infty d k \, k \,
e^{ - \frac{k^2}{2 C T}} \,  \sinh (2 \pi k)  \\
&\=  \frac{e^{S_0}\, \omega  \,
\coth \bigl(\pi \sqrt{2 C\omega}\, \bigr) }{Z(T) \, \pi} \, \sqrt{2  \pi} \, (C T)^{3/2} \, e^{2 \pi ^2 C T} \\
&\= 2 \,\omega \,  \coth \bigl(\pi \sqrt{2 C\omega}\, \bigr) \,.
 \end{split}
\ee
Upon further dropping~$(1+\tO(e^{-\omega/T}))$, we have
\be \label{eq:Tlomquant}
\text{Im} \, \CG^{\Delta=1}_{_R} (\omega)   \=
  \omega \,  \coth \bigl(\pi \sqrt{2 C\omega}\, \bigr) \,,
  \qquad \qquad \qquad \qquad \qquad \qquad \qquad \qquad \qquad
\ee
\begin{subnumcases}{\qquad \quad \=}
\omega \bigl(1+ \tO\bigl(e^{- 2 \pi \sqrt{2 C\omega}} \bigr) \bigr)  \,,  & $C\omega \gg 1$ \,, \label{eq:Tlomquanta} \\
 \dfrac{\omega}{\sqrt{2\pi^2 C\omega}} \Bigl(1+\dfrac{2\pi^2}{3} \,
 C\omega + \dfrac{4\pi^4}{45} \, (C\omega)^2 + \dots  \Bigr) \,,  & $C\omega \ll 1$ \,.\label{eq:Tlomquantb}
\end{subnumcases}
These results are consistent with the~$T \to 0$ limit given in~\cite{Mertens:2017mtv} and with the same limit of the
SYK model in~\cite{Bagrets:2016cdf,Bagrets:2017pwq}.

Equation~\eqref{eq:Tlomquant} summarizes Regime~II completely.
The result of Regime~IIa (which always satisfies~$1 \ll C\omega$) is~\eqref{eq:Tlomquanta},
and the result of Regime~IIb (which always satisfies~$C\omega \ll 1$) is~\eqref{eq:Tlomquantb}.
The errors in the various cases are detailed above.

\smallskip

\subsection{Summary of results for Green's functions}

\smallskip

We have calculated the Green's functions and the imaginary part of the retarded Green's function
in the Schwarzian theory in different regimes of approximation.
Whenever there is overlap in the approximations, the results agree.
When~$T$ is the largest scale and is in the semiclassical regime, i.e.~$\omega, E_\text{gap} \ll T$,
the results are given in~\eqref{eq:GomlargeT}--\eqref{eq:ImGintklargeT}.
When~$\omega$ is the smallest scale, and in the quantum regime i.e.~$\omega \ll T,  E_\text{gap}$,
the results are given in~\eqref{eq:Gomsmallom3}--\eqref{eq:omlTquant2}.
When~$T$ is smaller than~$\omega$,
the results are given in~\eqref{eq:GomsmallT}--\eqref{eq:Tlomquant} for a large range of~$T$.

\smallskip

We have verified all our analytical approximations by numerically computing the integrals.
In each case, we obtain very good agreement between our analytical approximations
and the numerical results over a range of values.
For example, we have checked that the ratio of value of the integral~\eqref{eq:Gintk1} and the value of
the formula~\eqref{eq:Gomsmallom3} is~1 to four decimal digits for the values~$C=1$, $\omega=10^{-8}$
over the range~$T \in [10^{-4},10]$. (The accuracy is the smallest when~$\omega=10^{-4}$, in which
case the ratio equals~$1.000049\dots$.)

\smallskip

Whenever at least one of the background or the scattered wave is in the semiclassical regime,
the imaginary part of the retarded Green's function agrees with the semiclassical hydrodynamic result,
i.e.~$\text{Im} \, \CG^{\Delta=1}_{_R} (\omega) = \omega$.
Note that, since in this case, one of~$T$ or~$\omega$ could be below~$E_\text{gap}$,
this is still a non-trivial statement about the quantum theory.
Finally---and most importantly---when both the background and the scattered wave are in the quantum regime
(i.e.~$CT$, $C\omega \ll 1$),
the second lines of~\eqref{eq:Gomsmallom3}, \eqref{eq:Tlomquant} show a strong deviation from the semiclassical formula.
In particular, the semiclassical result for the imaginary part of the Green's function is enhanced
by~$1/\sqrt{CT}$ when~$\omega \ll T$ and~$1/\sqrt{C\omega}$ when~$T \ll \omega$.

\section{Quantum shear viscosity and~$\eta/s$ \label{sec:queta}}

In this section we use the results on the Green's function in the previous section, in order to
extract the quantum value of the shear viscosity as a particular limit.
We first assemble the main idea that we discussed in the previous sections.
The two-point function in the boundary 3d theory starts as a Witten diagram near the boundary of AdS$_4$.
The two propagators can be followed all the way to the deep IR region, where they
couple to operators on the boundary of AdS$_2$. The UV Green's function gets a contribution from the propagators
as well as the IR two-point function in the AdS$_2$ theory.
In the semi-classical approximation, this IR two-point function is determined by propagators inside the classical
AdS$_2$ geometry. In the quantum theory, we have seen that this IR two-point function should be replaced by the
path integral calculation that we described in Section~\ref{sec:quGreen}.

\begin{figure}[h!]
        \centering
        \includegraphics[scale=1.2]{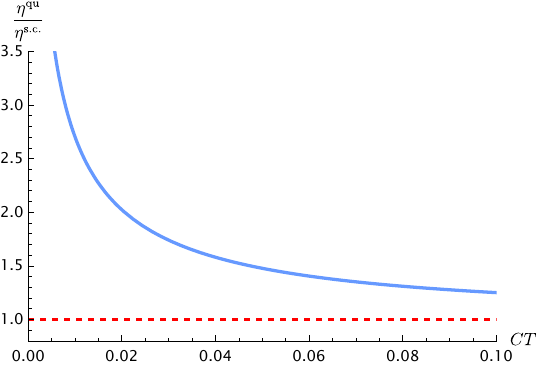}
    \caption{Plot of  $\eta^\text{qu}/\eta^\text{s.c.}$  calculated in the exact Schwarzian theory with~$\mu=10$.
    At large values of~$CT$ this reaches the semiclassical limit of~1, while at small~$CT$ there is a
    divergence of the form~$1/\sqrt{CT}$. This result can be trusted at the lower end when~$CT$ is much above
    the non-perturbative scale~$\approx e^{-\frac23 S_0} \approx 10^{-29}$.}
    \label{fig:ImGres}
\end{figure}

Upon putting all this together, we obtain the following formula for the quantum shear viscosity,
\be
\eta^\text{qu}
\= \frac{r_+^2}{16 \pi G} \,\lim_{\omega \to 0} \, \frac{\text{Im} \,  \CG^{\Delta=1}_{_R}(\omega)}{\omega}
\= \eta^\text{s.c.}\lim_{\omega \to 0} \, \frac{\text{Im} \,  \CG^{\Delta=1}_{_R}(\omega)}{\omega}\ .
\ee
The temperature  affects the result in two ways, as mentioned at the end of Section~\ref{sec:SCGreen}.
Firstly, there is a semi-classical correction to the zero-temperature formula which arises from the fact that the horizon
has radius~$r_+ = \rext + 2 \pi T L^2/3$ as given in~\eqref{eq:rMQT}.
Including this effect gives
\be
\eta^\text{s.c.} (\mu, T) \= \frac{1}{4\pi} \, s^\text{s.c.} (\mu, T)
\ee
with~$s^\text{s.c.} (\mu, T) $ given in~\eqref{eq:squ}.
Secondly, we need to take into account the quantum fluctuations of the AdS$_2$ region
in the calculation of the Green's function.
The expression~\eqref{eq:Gomsmallom3} leads to the following limiting expression,
\be \label{eq:ImGres}
\frac{\eta^\text{qu}}{\eta^\text{s.c.}} \=
\lim_{\omega \to 0} \, \frac{\text{Im} \,  \CG^{\Delta=1}_{_R}(\omega)}{ \omega}
\=
\text{erf}\left(\sqrt{2\pi^2 \,C T} \right) \Bigl( 1 + \frac{1}{4 \pi ^2 C T}\Bigr)
+ \frac{e^{-2 \pi ^2 C T}}{\sqrt{2 \pi^3 \, C T} }  \,.
\ee
In the two limits of small and large~$CT$, the limit \eqref{eq:ImGres} behaves as
\be \label{eq:ImGreslims}
\frac{\eta^\text{qu}}{\eta^\text{s.c.}} \=
\= \begin{cases} \vspace{0.2cm}
1 + \dfrac{1}{4 \pi ^2 CT} + \tO \bigl( e^{-CT} \bigr)  \, ,
  & \quad CT \gg 1 \,, \\
\Bigl(\dfrac{2}{\pi^3 CT} \Bigr)^\frac12 \,
 \biggl(1 +\dfrac{2 \pi^2 }{3} \, C T - \dfrac{2 \pi^4}{15} \, (C T)^2 + \dots \biggr)  \,,
& \quad CT \ll 1 \,.
\end{cases}
\ee
We observe that for~$CT \gg 1$, the first line above reaches the constant semi-classical limit
asymptotically.
On the other hand,
there is a drastic modification in the quantum regime~$CT \ll 1$ compared to the semi-classical one,
with a growth of~$(CT)^{-1/2}$ towards low temperatures.
The plot of the exact formula~\eqref{eq:ImGres} as a function of~$CT$ is given in~Figure~\ref{fig:ImGres}.
The recent papers~\cite{Brown:2024ajk,Emparan:2025sao,Biggs:2025nzs,Emparan:2025qqf,Betzios:2025sct}
calculate the quantum scattering cross-sections
in the Hamiltonian formalism of scattering by taking, as an input, the density of states in the Schwarzian theory,
and a coupling of the black hole degrees of freedom with the external fields.
Here we use the functional integral formalism to calculate the related Green's function.
It is gratifying to observe  that our results for the residue of the Green's function
are completely consistent with the scattering cross section  at fixed temperature~\cite{Biggs:2025nzs},
as expected on general grounds.

\bigskip

\begin{figure}[h!]
        \centering
        \includegraphics[scale=1.2]{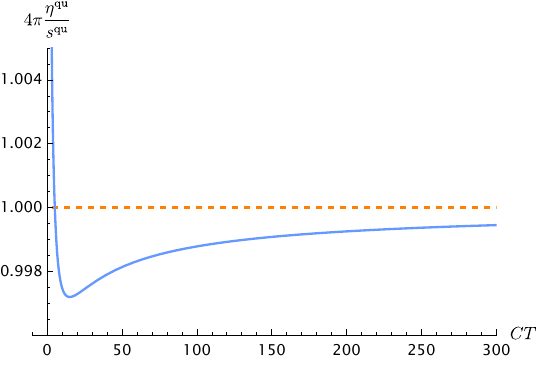}
    \caption{Plot of  $4 \pi \, \eta^\text{qu}(CT)/s^\text{qu}(CT)$   calculated in the exact Schwarzian theory with~$\mu=10$.
     This ratio reaches the value~1 asymptotically as~$CT\gg 1$. There is a minimum value
     at~$CT \approx 15$.
     At very small values of~$T/E_\text{gap}$ (but still larger than the non-perturbative scale~$\approx e^{-\frac23 S_0} \approx 10^{-29}$),
     the curve has a divergence of the form~$\sqrt{E_\text{gap}/T}$.}
    \label{fig:EtaovsPlot}
\end{figure}
\ndt Now we have all the ingredients to calculate the ratio
\be
\frac{\eta^\text{qu}}{s^\text{qu}}
 \= \frac{\eta^\text{qu}/s^\text{s.c.}}{s^\text{qu}/s^\text{s.c.}}
\= \frac{1}{4\pi} \frac{\eta^\text{qu}/\eta^\text{s.c.}}{s^\text{qu}/s^\text{s.c.}} \,.
\ee
We have already calculated the numerator on the right-hand side in~\eqref{eq:ImGres} and
the denominator is in~\eqref{eq:squ}, so the final answer, plotted in Figure~\ref{fig:EtaovsPlot}, is
\be \label{eq:EtaovsAns}
\begin{split}
& \frac{\eta^\text{qu}}{s^\text{qu}} \= \frac{1}{4\pi}\biggl( \text{erf}\left(\sqrt{2\pi^2 \,C T} \right) \Bigl( 1 + \frac{1}{4 \pi ^2 C T}\Bigr)
+ \frac{e^{-2 \pi ^2 C T}}{\sqrt{2 \pi^3 \, C T} }  \biggr) \, \bigg{/} \,
\biggl( 1 + \frac{\log \bigl((C T)^{\frac32} /\sqrt{2 \pi} \,\bigr)}{S^\text{s.c.}(\muext,T)} \biggr) \,.
\end{split}
\ee

\ndt We observe  that the ratio~$\eta^\text{qu}/s^\text{qu}$ reaches the semi-classical value~$1/4\pi$ asymptotically as~$CT$ becomes large.
Moving towards smaller values of~$CT$, there is a dip towards a minimum value,
and then a divergent climb
as~$E_\text{gap}/T \to 0$.

This result gives some credibility  to the expectation that at sufficiently low-temperatures,
the dynamics may be glassy, as signaled by the very large viscosity to entropy ratio.
We shall add further comments on this in the next section.

\section{Outlook}\label{outlook}

\ndt Our results are a first step towards addressing a host of questions associated with quantum near-extremal dynamics.

\smallskip

\ndt $\bullet$ Understanding the nature of the low-energy dynamics in the  ``quantum" regimes is extremely interesting. We need to determine whether the quantum effects still preserve the semiclassicality of the description, and whether they provide some small quantum corrections. If this is the case, then a Langevin-type description may be possible and remains to be discovered. The alternative is strong quantum effects, so that quantum uncertainties are large. In that case, only correlators will serve as a proxy for the low-energy dynamics.

\smallskip

There is a related paradigm of such a case, and this the case of baryons in a large N gauge theory, \cite{Witten}.
In this case, the semiclassical baryon is a soliton of the effective field theory (the {non-linear} $\s$-model with the WZ term).
However, the soliton carries quantum degrees of freedom, that generate the spin and isospin quantum numbers.
The dynamical description is a hybrid between the semiclassical properties of the soliton and the dynamics of
the quantum degrees of freedom.
This presents an example of a realization of the first possibility mentioned above, but has also
differences with the problem at hand.

\smallskip

\ndt $\bullet$ The Green functions for $\omega\gg E_{gap}$ are also very interesting and intriguing.
A central question is whether such Green functions can receive important quantum corrections.
Our results for the IR correlator suggest that quantum corrections in this regime are subleading. This includes cases Ia and IIa  in section \ref{regime}.
It would be interesting, however, to understand if a non-trivial momentum dependence may alter such a result, although we do not think so.

\smallskip

\ndt $\bullet$ There are other transport coefficients (IR limits of two-point functions in the massless sector (conserved charges). They can be computed similarly to what was done in this paper. There is also a further issue. There are relations between IR limits of two-point functions of the energy momentum tensor and the current. Some are also related to thermodynamic susceptibilities. The interesting question is what happens to such relations after including the quantum corrections.

\smallskip

\ndt $\bullet$ A dynamical instability of extremal black holes in the classical theory was discovered by the mathematician Aretakis \cite{Aretakis:2011ha, Aretakis:2012ei}. In the simplest case, a massless scalar field fluctuation of asymptotically flat extreme Reissner–Nordström (RN), decays everywhere on and outside the horizon but has radial derivatives which grow without bound at late times on the extreme horizon.
Subsequent analytic and numerical work in the mathematics and physics literature has established the Aretakis instability as a robust phenomenon applicable to a variety of perturbing fields, including massive scalars and (coupled) gravitational and electromagnetic perturbations, on extreme backgrounds of varying dimensions and asymptotics (see e.g. \cite{Lucietti:2012xr, Lucietti:2012sf, Murata:2012ct}).

The Aretakis behavior can be seen in the AdS$_2\times $S$^2$ near-horizon geometry of extreme RN and is intimately related to the symmetries of AdS$_2$ \cite{Lucietti:2012xr, Hadar:2018izi}.
In \cite{Gralla}, in the context of a near-extreme asymptotically AdS black hole, the Aretakis instability was connected to the behavior of correlators in the non-standard regime, $T\ll \omega,q\ll \mu$. In particular, it was shown, that correlators in this regime show the one-dimensional scaling (in $t$) expected from a one-dimensional CFT. In the limit $T=0$, this becomes the exact IR scaling of the IR CFT$_1$.
It is interesting to study the fate of the Aretakis instability in the quantum regime.

\vskip 0.5cm

\ndt $\bullet$ There are intriguing similarities between the SYK model behavior and classical glassy behavior that were expanded upon in \cite{Kurchan}.
In particular, the analogue of the glass transition in the disordered models happens at $T=0$ in the SYK Model. Just above $T=0$, the dynamics becomes slow and as $T\to 0$, the equilibration time in out-of-equilibrium correlation functions diverges. There is an analogue of the emergent one-dimensional  scale invariance in glassy systems (but the full SL(2,R) is absent).
The analogue of the Schwarzian modes corresponds to the correlated motion of larger and larger chunks of the system.
In this regime, the system develops an increased sensitivity to shear deformations. This rhymes constructively with the enhanced value of $\eta$, that we find in this work.
There is also a  difference between our case and classical glasses. Here the Schwarzian modes generate quantum dynamics whereas in glasses they fluctuate thermally.

In glasses, a semiclassical description with Langevin noise can describe the dynamics. This is an extra reason to believe that this will be the case for our system.
A further observable that is crucial for the glass transition is the behavior of the four-point function of the Schwarzian modes,
as it controls the fluctuations of the order parameter (which is  the two-point function).
This is important to calculate and verify indeed the aforementioned criticality.
Another important (but difficult) observable to calculate would be two-point functions out of equilibrium, in order to track their approach to equilibrium.

Our results on the viscosity to entropy ratio diverging at low temperatures is in agreement with the expectation that the low-energy dynamics of the system becomes glassy. This is also corroborated by the existence
of a large number of states at very low temperatures.

The picture of near extremal dynamics as nearly glassy dynamics rhymes interestingly with \cite{Anninos:2011vn}, where complex near-extremal multi-center black holes were constructed, that exhibited glassy dynamics.
It is also intriguing, whether there is also a correspondence with the recently studied ``grey galaxies" that fill the parameter space of black holes in $\mathcal{N}=4$ SYM, \cite{Choi:2024xnv}.
We plan to investigate these questions in the future.

$\bullet$ Our final comment refers to the 3rd law and its classical realization. It is known through recent work, \cite{KU} that in the classical gravitational theory, the third law of black hole mechanics is not valid. In \cite{KU}, the authors constructed solutions starting from regular initial data, that become extremal in finite time. This cast a doubt on the validity of the third law, via holography to the dual theories.
Our results suggest that the quantum effects, omnipresent in sufficiently near-extremal black holes, will change this conclusion. Indeed, a large IR shear viscosity and the associated analogy with glasses, indicates that relaxation times would diverge  as $T\to 0$, restoring the validity of the third law of thermodynamics in the quantum regime.

\bigskip

\section*{Acknowledgements}
\addcontentsline{toc}{section}{Acknowledgments}

We thank  D. Anninos, M. Dafermos, R.~Emparan, G. Fournodaylos, B. Gouteraux, C.~Herzog, L.~Iliesiu, J. ~Kurchan,
V. Niarchos, F. Nitti, D. Ramirez, M.~Rangamani, C. Rosen, C. Supiot, G. Turiaci, M.~Usatyuk, and D.~Vegh for useful and enjoyable discussions.
We especially thank P. Betzios for his participation and contributions in the initial stages of this work.
We thank R.~Emparan and M.~Rangamani for comments on a draft of the paper.

S.M. acknowledges the support of the STFC UK grants ST/T000759/1, ST/X000753/1.
This work was partially supported by  the H.F.R.I. call “Basic research Financing (Horizontal support of all Sciences)
under the National Recovery and Resilience Plan “Greece 2.0” funded by the European Union --NextGenerationEU
(H.F.R.I. Project Number: 15384), by the In2p3 grant ``Extreme Dynamics", the ANR grant ``XtremeHolo"
(ANR project n.284452), by the H.F.R.I. Project Number: 23770  of the H.F.R.I call
``3rd Call for H.F.R.I.'s Research Projects to Support Faculty Members \& Researchers"
and the UoC grant number 12030.

%\newpage

\vspace{0.6cm}

\appendix
	
\begin{appendix}
\renewcommand{\theequation}{\thesection.\arabic{equation}}
\addcontentsline{toc}{section}{Appendices}
%\section*{APPENDIX}

\section*{APPENDICES}

\section{The wave equation in the near-extremal BH background \label{app:waveeqn}}

In this appendix, we discuss the scattering of waves off the brane in the semi-classical approximation.
The scattering of waves off the brane in the extremal limit has been studied in detail
in~\cite{Edalati:2009bi,Edalati:2010hk,Davison:2013bxa}, where the focus is on taking the singular~$T \to 0$ limit.
Since we are interested in the quantum features at low temperatures, we introduce a small temperature,
and work out carefully the scattering of waves to leading order in $T$.

Consider a massless, neutral scalar~$\v$ propagating in the black brane background \eqref{eq:BBmetric}, governed by the wave equation
\be
\Box \, \v  \= 0 \,.
\ee
The translational symmetries of the background allow us to expand the field
into eigenmodes of energy and momenta as follows,
\be
\v(t,r,x_1,x_2) \= \sum_{\vec{q}}\, e^{2 \pi iq_1 x_1+ 2 \pi i q_2 x_2}\,
\int_{-\infty}^\infty d\omega\, e^{-i \omega t}  \, R_{\vec{q},\omega}(r) \,.
\ee
The sum runs over the integer-valued transverse momenta $\vec{q}=(q_1,q_2)$.
The mode~$R_{\vec{q},\omega}(r)$ obeys the following radial equation
\be \label{eq:radialeqn}
\frac{d}{dr}\, \Bigl( r^2 \, f(r) \, \frac{d}{dr}  R(r)  \Bigr)
+ \Bigl( \, \frac{\omega^2\, r^2}{f(r)} - L^2 \, k^2 \Bigr) \, R(r) \= 0\,,
\ee
where $q^2 = q_1^2 + q_2^2$ and from now on we suppress the subscripts~$\vec{q},\omega$ on the radial wavefunction function~$R_{\vec{q},\omega}(r)$.
For the application that we are interested in this paper, i.e.~the transport coefficient, we take vanishing
transverse velocity, i.e.~$q^2=0$, which we impose from now on.

\medskip

In the near-extremal limit, $T \ll \mu$, the interesting physics of scattering is contained in a range of low
frequencies~$\omega \ll \mu$. In this appendix we remain agnostic with regards to the relative size
of $T/\mu$ and $\omega/\mu$, allowing for either $T\ll\omega$ or $\omega\ll T$ (as well as for $T \sim \omega$).

\bigskip

We use the new dimensionless radial coordinate
\be
z=\frac{r-r_+}{r_\star}\,,
\ee
which is related to the $\rho$ coordinate in \eqref{eq:deftaurho} by $z=\ve \rho$,
and choose units such that $r_\star=1$.\footnote{We  keep $L$ around though,
so $r_\star$ may then be restored by dimensional analysis.}
Recall that in the near-extremal limit~$\ve \ll 1$, $T/\mu$ can be traded for~$\ve$, cf.~\eqref{eq:Tverel}.
We shall solve the wave equation \eqref{eq:radialeqn}, to leading order
in  $T/\mu$ and $\omega/\mu$, using the method of matched asymptotic expansions.
For a rigorous application of the method, it is beneficial to also rewrite the wave equation in terms of a
rescaled field variable $Y(z)=R(z)/z$.
We can then write the radial wave equation \eqref{eq:radialeqn} as follows:
\begin{align}\label{Yeqn}
	& z^2  y(z)^2 (z+\ve )^2 Y''(z)
	+\left[z y(z) (z+\ve ) (z (z+\ve ) (4 r_++2 z-\ve )+y(z) (4 z+3 \ve ))\right] Y'(z)  \notag \\
	&+\left[ L^4 \omega ^2 (r_++z)^4+y(z) (z+\ve ) \left(z (z+\ve ) (4 r_++2 z-\ve )+y(z) (2 z+\ve )\right) \right] Y(z) =0\,,
\end{align}
where $y(z)=6 r_+^2+4 r_+ z-4 r_+ \ve +z^2-z \ve +\ve ^2$.

\medskip

We divide the spacetime outside the horizon into two regions, defined as follows,
\be \label{eq:NearFarRegions}
\text{Near-region:} \quad  z\; \ll \; 1 \,, \qquad
\text{Far-region:} \quad z\gg \textrm{max}(\ve,\omega) \,,
\ee
and solve Eq.\eqref{Yeqn} in each region separately. Then we match the solutions in the
\be \label{eq:OverlapRegion}
\text{Overlap-region:} \quad \textrm{max}(\ve,\omega) \ll z\; \ll \; 1 \,,
\ee
whose existence is guaranteed in the near-extremal, $\ve\ll 1$, low-frequency, $\omega\ll 1$, regime.

\bigskip

In the Near-region, Eq.~\eqref{Yeqn} reduces, to leading order, to
\begin{align}\label{YeqnNear}
	& 36 z^2 (z+\ve )^2 Y''(z)
	+36 z (z+\ve ) (4 z+3 \ve ) Y'(z)
	+\left[ L^4 \omega ^2+36 (z+\ve ) (2 z+\ve ) \right] Y(z) =0\,,
\end{align}
whose general solution is:
\be\label{YsolnNear}
Y^\textrm{near}(z)\=C^\textrm{near}_1 \, z^{-1}\left(\frac{z}{z+\ve }\right)^{\frac{i L^2 \omega }{6 \ve }}
+C^\textrm{near}_2 \, z^{-1}\left(\frac{z}{z+\ve }\right)^{\frac{-i L^2 \omega }{6 \ve }}\,.
\ee
In the Overlap-region the above reduces to:
\be\label{YGensolnNearOverlap}
Y^\textrm{near}(z\gg \ve) \=
\left(C^\textrm{near}_1 + C^\textrm{near}_2 \right)  z^{-1} + \left(C^\textrm{near}_2 - C^\textrm{near}_1 \right) \frac{i L^2 \omega }{6} z^{-2}  \,.
\ee
The boundary condition for an ingoing solution at the horizon, $z=0$, is that $C^\textrm{near}_1 =0$ and we observe that in this case the retarded AdS$_2$ Green's function~$\mathcal{G}_R$,
which is proportional to the ratio of the source and vev in \eqref{YGensolnNearOverlap}, is given by Equation \eqref{eq:AdS2Green} in the main text.

\bigskip

In the Far-region, Eq.~\eqref{Yeqn} reduces to
\begin{align}\label{YeqnFar}
	& z^2 \left(z^2+4 z+6\right)  Y''(z)
	+ 2 z \left(3 z^2+10 z+12\right) Y'(z)  +4 \left(z^2+3 z+3\right) Y(z) =0\,,
\end{align}
whose general solution is:
\be\label{YsolnFar}
Y^\textrm{far}(z)=C^\textrm{far}_1 \, z^{-1}
+C^\textrm{far}_2 \, z^{-2} \left[2 z \log \left(z^2+4 z+6\right)-4 z \log z +\sqrt{2} z \tan ^{-1}\left(\frac{z+2}{\sqrt{2}}\right)-6\right] \,.
\ee
In the Overlap-region the above reduces to:
\be\label{YsolnFarOverlap}
Y^\textrm{far}(z\ll 1)=\left( C^\textrm{far}_1+C^\textrm{far}_2 \left(2 \log 6 +\sqrt{2} \tan ^{-1}\sqrt{2}\right)   \right) z^{-1}
-6 \, C^\textrm{far}_2 \, z^{-2}\,.
\ee

\bigskip

Near the AdS$_4$ boundary, $z\to\infty$, the radial wavefunction $Y(z)$ is given by:
\be\label{eq:AdS4asympt}
Y^\textrm{far}(z\to\infty)=A \, z^{-1}+B \, z^{-4}\,,
\ee
with
\be \label{eq:AdS4asympt2}
A= \left(C^\textrm{far}_1+C^\textrm{far}_2\frac{\pi}{\sqrt{2}}\right)\,, \qquad B= -12 \, C^\textrm{far}_2 \,.
\ee
The retarded AdS$_4$ Green's function~$G_R$ is proportional to the ratio $B/A$ of the source and vev in \eqref{eq:AdS4asympt}, as obtained from a solution that is ingoing at the horizon, i.e. a solution $Y(z)$ subject to the boundary condition $C^\textrm{near}_1 =0$.
Matching the near and far solutions in the Overlap-region,
that is matching Eqs.~\eqref{YGensolnNearOverlap} and \eqref{YsolnFarOverlap}, with $C^\textrm{near}_1 =0$, we find:
\begin{align}
	&\frac{C^\textrm{far}_1}{C^\textrm{near}_2}=\frac{2 i L^2 \omega  \log 6+i \sqrt{2} L^2 \omega  \tan ^{-1}\sqrt{2}+36}{36} \,, \\
	&\frac{C^\textrm{far}_2}{C^\textrm{near}_2}=-\frac{i L^2 \omega }{36} \,.
\end{align}
Therefore, from \eqref{eq:AdS4asympt2}, upon reinstating~$\rext$, we obtain:
\be
{B\over A}={1\over 3\rext}\frac{i L^2 \omega}{1+i \xi L^2 \omega }\,,
\ee
where~$\xi=\frac{-\frac{\pi }{2}+\sqrt{2} \log 6+\tan ^{-1}\sqrt{2}}{18 \sqrt{2} \, \rext}$.
This result is used to derive Equation~\eqref{eq:GUVIRrel} in the main text.
Note that, to leading order in $\omega$ and $T$, the above ratio is temperature-independent.

\section{On the dimension of the IR operator\label{B}}

In the set-up discussed in this paper, we have AdS$_2$ space that arises as the IR limit of a holographic
RG flow starting from AdS$_4$. Most of the discussion concerns a massless field in this bulk geometry.
In the main text we have assumed that the dimension of the operator in the IR CFT$_1$
(denoted by~$\Delta^{\text{IR}}$ in this appendix)
that couples to the massless scalar field  is~$\Delta^{\text{IR}}=1$.
In this appendix we discuss and justify this in more detail.

Recall that the mass of a scalar field~$\phi$ in AdS$_{d+1}$ space is related to
the conformal dimension~$\Delta$ in the dual~CFT$_d$ as~$m^2 = \Delta(\Delta-d)$.
The two solutions are \hbox{$\Delta_\pm = \frac{d}{2} \pm \nu$}, with~$\nu = \sqrt{\frac{d^2}{4}+m^2}$
where we have set the AdS length scale to~$L=1$.

The corresponding behaviors of the field at the boundary is
\hbox{$\phi_1 \sim u^{\Delta_-}$}, \hbox{$\phi_2 \sim u^{\Delta_+}$}.

In the conventional quantization with Dirichlet-type boundary condition, the leading mode~$\phi_1$ is kept fixed
and the subleading mode is allowed to fluctuate. Correspondingly, the dimension of the operator that couples
to the dynamical field is~$\Delta_+$.
However, in the range of masses such that~$\nu \in (0,1)$, there is also an alternative quantization with
Neumann-type boundary condition that is consistent~\cite{Breitenlohner:1982jf}.
In this case, the subleading mode~$\phi_2$ is kept fixed and the leading mode fluctuates,
and the dimension of the operator that couples to the dynamical field is~$\Delta_-$.

A massless scalar in AdS$_2$ space falls precisely in this window, and
the two corresponding operator dimensions are~$\Delta=1$ and~$\Delta=0$.
Therefore, we are led to ask: what is the correct dimension of the operator in the CFT$_1$ dual to
the massless scalar in IR AdS$_2$ theory in our problem.

\medskip

We can determine~$\Delta^{\text{IR}}$ by performing the following small calculation.
Firstly, we impose Dirichlet-type boundary conditions in the UV theory,
i.e., we keep only the leading mode near the AdS$_4$ boundary.
In terms of the expression~\eqref{eq:AdS4asympt} near the AdS$_4$ boundary, this means that we have~$B=0$,
which, using~\eqref{eq:AdS4asympt2}, is equivalent to $C^\textrm{far}_2=0$.
With this boundary condition, the general far-solution~$Y^\text{far}$, given in~\eqref{YsolnFar},
reduces in the Overlap-region, using ~\eqref{YsolnFarOverlap}, to
\be\label{eq:YGenFarOverlap}
Y^\textrm{far}(z\ll 1) \=  C^\textrm{far}_1\, z^{-1} \,.
\ee

Given that this has to match to the near-solution in the Overlap-region,
Eqn.~\eqref{YGensolnNearOverlap}, we see that in AdS$_2$ only the leading-order mode is turned on at its boundary,
which implies that one has the conventional quantization and that
the dimension of the corresponding operator has IR dimension~$\Delta^{\text{IR}}=1$.

\smallskip

Note that the above matching is done only in the region between the AdS$_2$ boundary
(the Overlap-region) and the AdS$_4$ boundary and, in particular,  boundary conditions
at the horizon of the black hole are not imposed. This is consistent with the idea of holographic
RG flow.

\medskip

One can also reach the same conclusion in an equivalent, but slightly more systematic manner.
In order to do that, we recall that
the two types of boundary conditions in AdS space lead to two different dual CFTs at the
boundary~\cite{Klebanov:1999tb}.  The two CFTs dual to the two quantizations
are related by an RG flow triggered by a double trace operator~\cite{Witten:2001ua}.
Once we allow for deformations by this operator, both theories can be described in either quantization.

In the context of holographic RG flow, the papers~\cite{Heemskerk:2010hk,Faulkner:2010jy}
analyzed the question of what happens to a particular quantization in the AdS$_{d+1}$ theory as one flows to the~IR.
The analysis is done in the formalism of Neumann (alternate) quantization with the addition of a double-trace deformation.
The coupling of the double-trace operator in the action is proportional to the ratio $\chi = \beta/\alpha$, where $\beta$ is the
subleading  mode of the scalar field in the asymptotic region and $\alpha$ is the leading mode.

When~$\chi=0$, there is no deformation, and one has the alternate quantization.
When~$\chi=\infty$, one has reached the end-point of the RG flow triggered by the
deformation\footnote{To avoid confusion, the RG flow in this sentence (and in~\cite{Witten:2001ua})
is not the same as the holographic RG flow being discussed in this paper.
For example, in the language of~\cite{Witten:2001ua}, we think of conventional quantization at the
AdS$_4$ boundary already as the end-point of an RG flow, while for us it is the starting point.
Note also that, even if we ignore the interpretation of~\cite{Witten:2001ua} in the dual CFT,
the gravitational interpretation of~$\chi$ is clear, since the value of the field is the expectation
value of the operator: with~$\chi=0$ there is only a growing mode, and that is indeed the expectation value
in the alternate quantization; while with~$\chi=\infty$ there is only a subleading mode, and that is indeed
the expectation value in the conventional quantization.},
which is simply the conventional quantization.

We follow~\cite{Faulkner:2010jy}, in which, the method of analysis is as follows.
Firstly, one recalls that the quantization conditions are
controlled in the gravitational theory by the boundary action of the theory. Then one writes a flow equation
for this boundary action using the Hamilton-Jacobi equations of holographic RG flow.
These equations for the coupling in the boundary theory that controls the two-point function
can effectively be solved using a matching procedure.
As a result, one obtains a formula for~$\chi_{_\text{IR}}$ in terms of~$\chi_{_\text{UV}}$,
from which we can deduce the respective couplings and dimensions of operators.

In order to present this formula, we need to gather some data.
Let us denote the UV boundary by the radial coordinate~$u=0$ in the Fefferman-Graham expansion.
At the AdS$_4$ boundary, the leading and subleading modes of the scalar field behave
as
\be \label{eq:UVbehavior}
\phi_1 \; \sim \; u^{\Delta^\text{UV}_-} \,, \qquad \phi_2 \;\sim \; u^{\Delta^\text{UV}_+} \,,
\qquad u \to 0 \,,
\ee
respectively.

Near the AdS$_2$ boundary, which may be placed at an arbitrary surface $u=u_\ast$ in the Overlap-region from Appendix \ref{app:waveeqn}, we choose a basis of solutions which have the scaling behavior
governed by the dimensions in the IR theory as
\be \label{eq:IRbehavior}
\eta_\pm \; \sim \; (u - u_\ast)^{-\Delta^\text{IR}_\mp} \,, \qquad u \to u_\ast \,.
\ee

The two bases are matched by the following equations,
\be
\eta_\pm \= a_\pm \phi_1 + b_\pm \phi_2 \,.
\ee
The relation between the couplings in the IR and the UV theory is then governed by
\be
\chi_{_\text{IR}} \= - \frac{b_+ - \chi_{_\text{UV}} \, a_+}{b_- - \chi_{_\text{UV}} \, a_-} \,.
\ee

In our case, the AdS$_4$ boundary is at~$z=\infty$, and the UV behavior of the field
given in~\eqref{eq:AdS4asympt}. We can read off the UV dimensions  in the notation of~\eqref{eq:UVbehavior}
to be~$\Delta^\text{UV}_+=3$, $\Delta^\text{UV}_-=0$.
The horizon is at~$z=0$, and the IR behavior of the field near the AdS$_2$ boundary is given in~\eqref{YGensolnNearOverlap}
(and also in~ \eqref{YsolnFarOverlap}).
We can read off the IR dimensions in the notation of~\eqref{eq:IRbehavior}
to be~$\Delta^\text{IR}_+=1$, $\Delta^\text{IR}_-=0$.
The matching analysis is precisely the one discussed near~\eqref{eq:YGenFarOverlap},
which shows that~$a_-=0$, $a_+ \neq 0$.
Finally, at the AdS$_4$ boundary we impose conventional quantization, so that
we have~$\chi_{_\text{UV}}=\infty$. Upon putting these facts together, we obtain~$\chi_{_\text{IR}}=\infty$,
which means that we also have conventional quantization in the IR, with the corresponding
operator dimension~$\Delta^\text{IR}=\Delta^\text{IR}_+=1$.

\bigskip

\section{Details of calculations of the Green's function \label{app:intcalcs}}

In this appendix we present details of the calculations underlying the evaluation of the Green's function in
Section~\ref{sec:quGreen}, in the various regimes of approximation.
The starting point is the integral~\eqref{eq:Gintk1}, which we recall here,
\be \label{eq:Gintk1A}
\begin{split}
\CG^{\Delta=1} (\omega) &  \=
 \frac{e^{S_0}\, \omega}{Z(T) \, 2 \pi^2} \,  \int_0^\infty d k \; g_{_T}(k)
 \,\times f(k,\omega) \,,\\
\text{with} \qquad \qquad g_{_T}(k) &\; \coloneqq \; k \, e^{ - \frac{k^2}{2 C T}} \, \sinh (2 \pi k) \\
f(k,\omega) & \; \coloneqq \; \frac{\sinh \bigl(2 \pi \sqrt{k^2 + 2 C \omega}\, \bigr)}{\sinh \bigl(\pi(\sqrt{k^2 +  2C \omega}+k) \bigr)
\sinh \bigl(\pi(\sqrt{k^2 +  2C \omega}-k) \bigr)} \,. \qquad
\end{split}
\ee
The partition function~$Z(T)$ is given in~\eqref{eq:Zresult}.

\medskip

\begin{figure}[h!]
        \centering
        \includegraphics[scale=0.7]{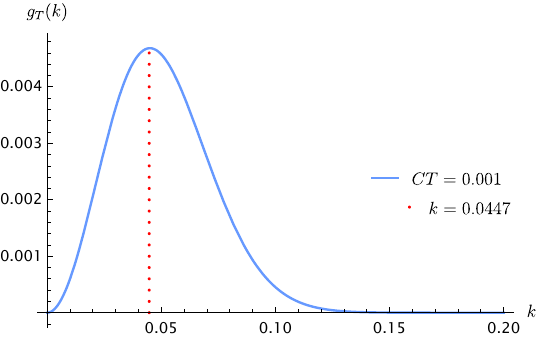}
        \includegraphics[scale=0.7]{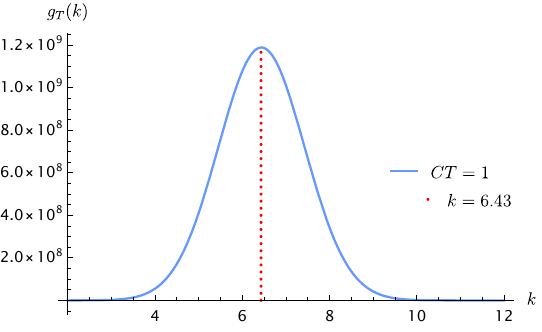}
        \includegraphics[scale=0.7]{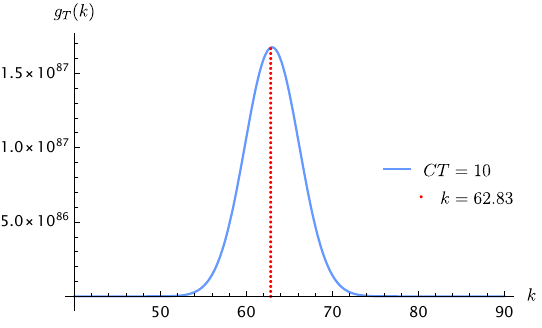}
    \caption{Plots of  $g_{_T}(k)$ calculated with~$CT=0.001, 1, 10$.
    Note that the ranges of the vertical axes in the three plots are different.
    In each case there is a bell-shaped region, but the maximum value increases rapidly with~$T$.}
    \label{fig:gPlot}
\end{figure}

The function~$g_{_T}(k)$ is shown in Figure~\ref{fig:gPlot} for different values of~$T$.
It vanishes at~$k=0$ and has the following behavior as~$k \to 0$,
\be
g_{_T}(k)\=2\pi k^2+\tO(k^4)   \,.
\ee
For large values of~$k$, it decays to zero exponentially as~$k \to \infty$.
It has a single maximum at $k=k_*$ with
\be
{k_*^2\over CT}-1 \=2\pi k_* \coth(2\pi k_*) \,.
\ee
For $CT\ll 1$ we have
\be
k_*=\sqrt{2CT} \, \bigl( 1+\tO(CT) \bigr) \,, \qquad
g_{_T}(k_*) \; \approx \; \sqrt{2CT}e^{-1}\sinh\bigl(2\pi\sqrt{2CT} \bigr) \; \approx \; 4\pi CT\,e^{-1} \,.
\ee
For $CT\gg 1$ we have
\be
k_*\=2\pi CT+{1\over 2\pi}+\tO \Bigl({1\over CT} \Bigr) \,, \qquad
g_{_T}(k_*) \; \approx \;  e^{2 \pi^2 CT}\,.
\ee
As we see in the figures, $g_T(k)$ has a bell-shaped region outside of which it is highly suppressed.
This region moves to the right as~$T$ increases.

\medskip

\begin{figure}[h!]
        \centering
        \includegraphics[scale=1]{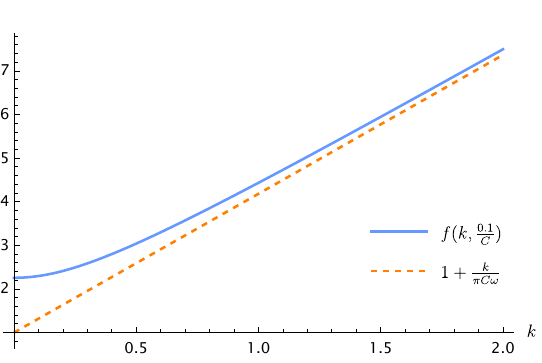}
    \caption{Plot of  $f(k,\omega)$ and~$1+ \frac{k}{\pi C\omega}$ calculated with~$C\omega=0.1$.
    $f$ is approximately constant for small values of~$k$, and $f$ is approximately a linear function at large values of~$k$.}
    \label{fig:fPlot}
\end{figure}

The function~$f$ is shown in Figure~\ref{fig:fPlot}.
It is regular as $k\to 0$ with
\be \label{eq:fsmallk}
f(0,\omega) \= 2  \coth \bigl(2 \pi \sqrt{2C\omega}\, \bigr) \,,
\ee
while it diverges linearly as $k \to \infty$,
\be \label{eq:flargek}
f(k,\omega) \=\frac{k}{\pi C\omega}+1 + \tO \Bigl(\, \frac{1}{k} \, \Bigr) \,.
\ee

\bigskip

We now evaluate the integral~\eqref{eq:Gintk1A}, in different regions of approximation of the
parameters~$\omega$ and~$T$.
As mentioned above, the function~$f$ transitions from a constant for small~$k$ to a linear behavior at large~$k$.
The basic intuition behind the approximation of the integral is to estimate the placement of the bell-shaped
region of~$g_T(k)$ with respect to the different regions of behavior of~$f(k)$.
When~$C\omega$ is parametrically smaller than~$CT$, then most of the bell region of the
function~$g(k)$ is situated in the linear region of~$f(k)$.
We denote this regime as Regime~{\bf I}. In this regime, we can approximate~$f$
by its leading linear behavior in~\eqref{eq:flargek}.
The integral~$\int_0^\infty d k \, g_{_T}(k) \times f(k,\omega)$ can then be approximated by
\be \label{eq:gfR1int}
\begin{split}
I(\omega,T) & \=  \int_0^\infty d k \, \frac{k^2}{\pi C\omega} \, \sinh (2 \pi k) \, e^{ - \frac{k^2}{2 C T}}  \\
&  \=   \frac{(2\pi C)^\frac32}{\omega}  \, T^{5/2} \, e^{2 \pi ^2 C T}
\biggl(  \text{erf}\left(\sqrt{2} \pi  \sqrt{C T} \right) \Bigl( 1 + \frac{1}{4 \pi ^2 C T}\Bigr)
+ \frac{e^{-2 \pi ^2 C T}}{\sqrt{2 \pi^3 \, C T}}  \biggr) \,. \\
\end{split}
\ee
On the other hand, when~$CT$ is parametrically smaller than~$C\omega$, then
most of the bell region of the function~$g(k)$ is situated in the constant region of~$f(k)$.
We denote this regime as Regime~{\bf II}. In this regime, we can approximate~$f$
by its leading constant behavior in~\eqref{eq:fsmallk}.
The integral~$\int_0^\infty d k \, g_{_T}(k) \times f(k,\omega)$ can then be approximated by
\be\label{eq:gfR2int}
\begin{split}
J(\omega,T) &  \approx  f(0,\omega)  \int_0^\infty d k \, k \, \sinh (2 \pi k) \, e^{ - \frac{k^2}{2 C T}}
 \=   f(0,\omega) \, \sqrt{2} \pi ^{3/2} e^{2 \pi ^2 C T} (C T)^{3/2} \,. \qquad
\end{split}
\ee

In both regimes, we can study the approximations to different orders of accuracy by
further considering the placement
of the parameters~$\omega$,~$T$, and~$E_\text{gap}$ or, equivalently, $C\omega$,~$CT$, and~1
with respect to each other.
We now discuss the details of these approximations.

\bigskip

\ndt {\bf Regime I.} \textit{$\omega$ is smaller than~$T$}\\
\ndt {\bf I a.} \textit{$1, C\omega \ll CT$}

\smallskip

In this regime of parameters, we express each of the three sinh functions in~$f$ in terms of
exponentials,
\be \label{eq:sinhexpapprox}
\begin{split}
f(k,\omega) & \= \frac{\sinh \bigl(2 \pi \sqrt{k^2 + 2 C \omega}\, \bigr)}{\sinh \bigl(\pi(\sqrt{k^2 +  2C \omega}+k) \bigr)
\sinh \bigl(\pi(\sqrt{k^2 +  2C \omega}-k) \bigr)} \\
& \= \frac{\bigl(1- e^{-4 \pi \sqrt{k^2 + 2 C \omega}} \, \bigr) }{\bigl(1-e^{-2\pi(\sqrt{k^2 +  2C \omega}+k)} \bigr)} \times
\frac{2}{\bigl(1-e^{-2\pi(\sqrt{k^2 +  2C \omega}-k)} \bigr)}
\end{split}
\ee
Since~$CT \gg 1$, $g_{_T}$ is peaked near~$k_*=2 \pi CT$.
In most of the bell-shaped region of~$g_T(k)$, $k \gg 1$ and so we can approximate the first factor by~1,
making an error of~$\tO(e^{-4 \pi k})$.
For the second factor we use
\be
\frac{2}{1-e^{-2\pi (\sqrt{k^2+2C\omega} -k)}} \= \frac{k}{\pi C \omega} \Bigl( 1+ \tO\Bigl(\frac{C\omega}{k} \Bigr) \Bigr) \,.
\ee
Upon putting these together, we obtain, in this regime,
\be
f(k,\omega) \= \frac{k}{\pi C \omega} \Bigl( 1+ \tO\Bigl(\text{max} \Bigl(\frac{C\omega}{k}  , e^{-4 \pi k} \Bigr) \Bigr) \,.
\ee

Since~$g_T(k)$ is approximated by a Gaussian centered at~$k=2\pi CT$ with variance~$CT$,
we can estimate the error in the integral by setting~$k = \frac12 \pi CT$.
\footnote{One can set~$k = \alpha CT$, where~$\alpha$ is an~$\tO(1)$ real number, and
one can present a sharp bound for~$\alpha$ by consider a certain percentage of the bell-shaped region.}
Therefore, we obtain,
\be \label{eq:GomlargeTA}
\begin{split}
& \text{with an error of $\tO\Bigl(\text{max} \Bigl( \frac{2C\omega}{CT} \, , \, e^{- 2 \pi^2 C T} \, \Bigr) \Bigr) $} ,  \\
& \CG^{\Delta=1} (\omega) \\
& \quad  \= \frac{e^{S_0}}{Z(T) \, 2 \pi^3 C} \,  \int_0^\infty d k \, k^2 \, \sinh (2 \pi k) \, e^{ - \frac{k^2}{2 C T}}  \\
& \quad \=  \frac{e^{S_0}}{Z(T) \, 2 \pi^3 C} \, 2^\frac32  \, (\pi CT)^{5/2} \, e^{2 \pi ^2 C T}
\biggl(  \text{erf}\left(\sqrt{2} \pi  \sqrt{C T} \right) \Bigl( 1 + \frac{1}{4 \pi ^2 C T}\Bigr)
+ \frac{e^{-2 \pi ^2 C T}}{\sqrt{2 \pi^3 \, C T}}  \biggr) \\
&  \quad \=    2 \, T  \biggl(
\text{erf}\left(\sqrt{2} \pi  \sqrt{C T} \right) \Bigl( 1 + \frac{1}{4 \pi ^2 C T}\Bigr)
+ \frac{e^{-2 \pi ^2 C T}}{\sqrt{2 \pi^3 \, C T}}  \biggr) \\
&  \quad \=    2 \, T \, \Bigl( 1 + \frac{1}{4 \pi ^2 C T}\Bigr) \,.
 \end{split}
\ee
Here, we have used the result~\eqref{eq:gfR1int}, and then only kept terms consistent with the error.
Within the same approximation, we have
\be \label{eq:ImGintklargeTA}
\text{Im} \, \CG^{\Delta=1}_{_R} (\omega)  \= \frac12 \bigl( 1-e^{- \omega/T} \bigr) \, \CG^{\Delta=1} (\omega)
\= \frac{\omega}{2T} \, 2T \Bigl( 1 + \frac{1}{4 \pi ^2 C T}\Bigr)  \= \omega \Bigl( 1 + \frac{1}{4 \pi ^2 C T}\Bigr) \,.
\ee

\bigskip

\ndt {\bf I b.} \textit{$C\omega \ll 1, CT, (CT)^2$}

\smallskip

In this regime we use the fact that~$C\omega$ is the smallest scale,
and we allow~$CT$ to be either small or large compared to~1.
Since~$C\omega$ and~$CT$ can be much smaller than~1, we cannot approximate
any of the $\sinh$ functions with exponentials as in~\eqref{eq:sinhexpapprox}.
Instead, we proceed as follows.
First we estimate the function~$f$ for large~$k$ and show that
it is well-approximated by the linear function~$\frac{k}{C\omega}$. More precisely, writing
\be
f (k,\omega) \= f_1(k,\omega) \, f_2(k,\omega) \, \frac{k}{C\omega} \,,
\ee
with
\be
\begin{split}
f_1(k,\omega) \; \coloneqq \;  \frac{\sinh \bigl(2 \pi \sqrt{k^2 + 2 C \omega}\, \bigr)}{\sinh \bigl(\pi(\sqrt{k^2 +  2C \omega}+k) \bigr)} \,, \qquad
f_2(k,\omega)  \; \coloneqq \; \frac{\pi C\omega/k}{\sinh \bigl(\pi(\sqrt{k^2 +  2C \omega}-k) \bigr)} \,,
\end{split}
\ee
we show that both~$f_1$ and~$f_2$ are well-approximated by~$1$ in a region excluding a small region near~$k=0$.
Then we show that, in the regime of parameters chosen above,
the difference of the original integral~\eqref{eq:Gintk1A} and the integral with the replacement~$f \to \frac{k}{C\omega}$
in that small region is small compared to the value of the integral.

\medskip

\begin{figure}[h!]
        \centering
        \includegraphics[scale=1]{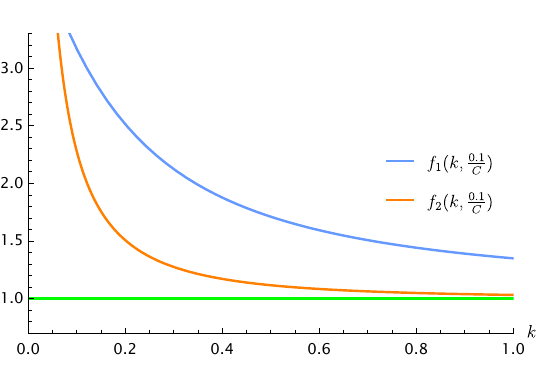}
    \caption{Plot of  $f_1(k,\omega)$ and~$f_2(k,\omega)$ calculated with~$C\omega=0.1$.
    $f_1$ and~$f_2$ are both decreasing functions, and reach~$1$ asymptotically as~$k \to \infty$.}
    \label{fig:f1f2Plot}
\end{figure}

To make the estimates, we first note that~$f_1$ and~$f_2$ are both decreasing functions (see Figure~\ref{fig:f1f2Plot})
that asymptotically reach the value~1 as~$k \to \infty$.

\medskip

For~$C\omega, k \ll 1$, the argument of the sinh functions are small
and we can approximate them by polynomials ($\sinh x = x + \frac16 x^3 + \dots$).
In the regime~$k \to 0$ and~$\xi=\frac{C\omega}{k^2} \to 0$, considering them as independent variables,
one obtains
\be
\begin{split}
f_1(k,\omega) & \= 1 + \sum_{n_1,n_2 > 0} \, a_1(n_1,n_2) \, (\pi k)^{2n_1} \, \xi^{n_2} \\
& \= 1+\frac{1}{2} \left(\xi-\xi^2 + \dots\right)
+ \Bigl(\,\frac{2  \xi}{3} + \frac{\xi^2}{2} + \dots \Bigr) (\pi k)^2
+\tO \bigl((\pi k)^4 \bigr) \,,
\end{split}
\ee
\be
\begin{split}
f_2(k,\omega) & \= 1 + \sum_{n_1,n_2 > 0} \, a_2(n_1,n_2) \, (\pi k)^{2n_1} \, \xi^{n_2} \qquad \\
& \= 1+\frac{\xi}{2}-\Bigl(\,\frac{1}{4}+\frac{1}{6} \pi ^2 k^2 \Bigr) \xi^2+
\Bigl(\,\frac{1}{4} + \frac{1}{12} \pi ^2 k^2 \Bigr) \xi^3+\tO \bigl(\xi^4 \bigr) \,.\qquad
\end{split}
\ee
Both functions are regular in this limit, and therefore there is no ambiguity in the series expansions.
Together, we have, as~$C\omega \to 0$, $k \to 0$, and~$\frac{C\omega}{k^2} \to 0$,
\be \label{eq:fapproxA}
f_1(k,\omega)\, f_2(k,\omega)
  \=  1+ \tO\Bigl( \text{max} \Bigl(\frac{C\omega}{k^2} ,  k^2  \Bigr)  \Bigr) \,.
\ee

\medskip

Now we split the integration region into two parts: the first part runs from~$k=0$ to~$k=k_0=(C\omega)^{1/4}$,  the second
part runs from~$k_0$ to~$k=\infty$.
\footnote{Since we are considering asymptotic estimates, we can assume that~$(C\omega)^\frac14$ can be made arbitrarily
small. For numerical approximations, we can put bounds on~$C\omega$ according to the required order of accuracy.  }
The equation~\eqref{eq:fapproxA} and the fact that~$f_1$, $f_2$ are decreasing together lead to the following estimate,
\be
f_1\bigl( k,\omega \bigr) f_2\bigl( k,\omega \bigr)  \= 1+  \tO\bigl( k_0^2 \bigr)  \,, \quad k \ge k_0 = (C\omega)^\frac14 \,,
\ee
and, consequently,
\be \label{eq:festlargek}
f\bigl( k,\omega \bigr) = \frac{k}{\pi C \omega} \bigl(1+ \tO\bigl( k_0^2 \bigr)   \bigr) \,, \quad k \ge k_0  \,.
\ee
In other words, we can approximate~$f$ by~$\frac{k}{\pi C \omega}$ in
the second part of the integration region, with a relative error~$\sqrt{C\omega}$.
Now we consider the first part of the integration region. Firstly, we note that the difference in the integrands, i.e.,
\be
h(k,\omega) \; \coloneqq \;
 g(k,\omega) \, \delta f(k,\omega) \= k \,e^{ - \frac{k^2}{2 C T}} \, \sinh (2 \pi k) \, \Bigl( f(k,\omega) -\frac{k}{\pi C \omega} \Bigr)
\ee
is an increasing function and vanishes at~$0$.
Hence the integral of~$h$ in the first part of the integration region is bounded above by~$k_0 \, h(k_0,\omega)$.
We have
\be \label{eq:firsrreg}
\int_0^{k_0} h(k,\omega) \; \le \; k_0 \, h(k_0,\omega) \= \frac{2 k_0^4}{k_0^4} \, \tO\bigl( k_0^2 \bigr)   \=  \tO\bigl( k_0^2 \bigr)  \,.
\ee

\bigskip

Therefore, we observe  that by dropping terms of~$\tO\bigl(\sqrt{C\omega}\bigr)$,
we can approximate~$f$ by~$\frac{k}{C\omega}$ for all~$k$.
With this approximation, the two-point function takes the following form,
using~$I(\omega,T)$ in~\eqref{eq:gfR1int},
\be \label{eq:Gomsmallom3A}
\begin{split}
 \CG^{\Delta=1} (\omega)
 & \=    2 \, T  \biggl(\text{erf}\left(\sqrt{2} \pi  \sqrt{C T} \right) \Bigl( 1 + \frac{1}{4 \pi ^2 C T}\Bigr)
+ \frac{e^{-2 \pi ^2 C T}}{\sqrt{2 \pi^3 \, C T}}  \biggr) \\
&  \= \begin{cases}
\quad 2T \Bigl(  1 + \dfrac{1}{4 \pi ^2 CT} + \tO \bigl( e^{-CT} \bigr)  \Bigr) \,, & \quad 1 \ll CT \,, \\[10pt]
 \dfrac{\sqrt{8 T} }{\sqrt{\pi^3 C}}  \,
\biggl(1 +\dfrac{2 \pi^2 }{3}  \, C T - \dfrac{2 \pi^4}{15} \, (C T)^2 + \dots \biggr)  \,, & \quad CT \ll 1 \,.
\end{cases}
 \end{split}
\ee
With the same approximation,
\be  \label{eq:omlTquant2A}
\begin{split}
\text{Im} \, \CG^{\Delta=1}_{_R} (\omega)  & \=
  \omega  \biggl(  \text{erf}\left(\sqrt{2} \pi  \sqrt{C T} \right) \Bigl( 1 + \frac{1}{4 \pi ^2 C T}\Bigr)
+ \frac{e^{-2 \pi ^2 C T}}{\sqrt{2 \pi^3 \, C T}}  \biggr) \,\qquad \\[4pt]
& \= \begin{cases}
\quad \omega \Bigl(  1 + \dfrac{1}{4 \pi ^2 CT} + \tO \bigl( e^{-CT} \bigr)  \Bigr) \,, & \quad 1 \ll CT \,, \\[10pt]
 \dfrac{\sqrt{2} \, \omega}{\sqrt{\pi^3 C T}}  \,
\biggl(1 +\dfrac{2 \pi^2 }{3}  \, C T - \dfrac{2 \pi^4}{15} \, (C T)^2 + \dots \biggr)  \,, & \quad CT \ll 1 \,.
\end{cases}
\end{split}
\ee

\bigskip

\bigskip

To estimate the error, we demand that the integral in the first part that we dropped after~\eqref{eq:firsrreg}
(bounded by~$\tO(\sqrt{C\omega})$)
does not overwhelm the value of the integral~$I(\omega,T)$ given in~\eqref{eq:gfR1int}.
For large~$T$, the relative error is therefore~$\tO(\sqrt{C\omega} \, e^{-2 \pi^2 CT})$.
For small~$T$, the relative error is larger and is estimated by~$\tO \Bigl(\dfrac{(C\omega)^\frac32}{(CT)^2} \Bigr)$,
which is small in the regime that is assumed here.
In addition, we have the relative error~$\tO \bigl( \sqrt{C\omega}  \bigr)$ dropped after~\eqref{eq:festlargek},
The overall relative error is therefore~$\tO \Bigl(\text{max} \Bigl(\sqrt{C\omega} , \dfrac{(C\omega)^\frac32}{(CT)^2} \Bigr)\Bigr)$.

\bigskip

\ndt {\bf Regime II.} \textit{$T$ is smaller than~$\omega$}

\smallskip

In this region, $C\omega$ is much larger than the peak of the integrand in~\eqref{eq:Gintk1A}.
Recall that the peak is situated around~$2 \pi CT$ for~$CT \gg 1$, and around~$\sqrt{CT}$ for~$CT \ll 1$.
So we seek to approximate the function~$f$ in~\eqref{eq:Gintk1A}
in the regime~$k \ll C\omega$.
The following identity is useful,
\be
f(k,\omega)  \= f_+(k,\omega) +f_-(k,\omega) \,,
\ee
with
\be
 f_\pm(k,\omega) \= \coth \bigl(\pi(\sqrt{k^2 +  2C \omega}\pm k) \bigr) \,.
\ee
These functions obey
\be \label{eq:fpmomls}
\begin{split}
\frac{f_\pm(k,C\omega)}{\coth \bigl(\pi\sqrt{2C \omega} \bigr) }
& \=  1 \mp  \frac{2\pi k}{\sinh \bigl(2 \pi \sqrt{2C\omega} \bigr)} + \dots  \\
& \= \begin{cases}
1 + \tO\bigl( \frac{k}{\sqrt{C\omega}} \bigr) \,,  \qquad  & C\omega \ll 1 \,, \\
1 + \tO\bigl( k \, e^{-\sqrt{C\omega}} \bigr) \,,  \qquad  & C\omega \gg 1 \,.
\end{cases}
\end{split}
\ee
It is clear that, as long as~$k \ll \sqrt{C\omega}$, the right-hand side is well-approximated by~1.

\bigskip

\ndt {\bf II a.} \textit{$1, CT, (CT)^2 \ll C\omega$}: The error can be estimated by setting~$k$ to be
near the peak of $g_T$, and depends on whether~$CT \gg 1$ or~$CT \ll 1$.
Using the second expression in~\eqref{eq:fpmomls}, it is given by
 $\tO\bigl( \text{max} \bigl(\sqrt{CT} \, e^{-\sqrt{C\omega}}\,, \, CT \, e^{-\sqrt{C\omega}} \bigr)\bigr)$.\\

\ndt {\bf II b.} \textit{$ CT \ll C\omega \ll 1$}: In this regime, $k$ can be set to be of~$\tO(\sqrt{CT})$, and
the error is given the first expression in~\eqref{eq:fpmomls} to be
 $\tO\bigl( \sqrt{\frac{CT}{C\omega}} \bigr)$. \\

With these errors, the two-point function takes the following form,
\be \label{eq:GomsmallTA}
\begin{split}
\CG^{\Delta=1} (\omega)
& \= \frac{e^{S_0}\, \omega \,  \coth \bigl(\pi \sqrt{2 C\omega} \,\bigr)}{Z(T) \,  \pi^2} \,  \int_0^\infty d k \, k \,
e^{ - \frac{k^2}{2 C T}} \,  \sinh (2 \pi k)  \\
&\=  \frac{e^{S_0}\, \omega  \,
\coth \bigl(\pi \sqrt{2 C\omega}\, \bigr) }{Z(T) \, \pi} \, \sqrt{2  \pi} \, (C T)^{3/2} \, e^{2 \pi ^2 C T} \\
&\= 2 \,\omega \,  \coth \bigl(\pi \sqrt{2 C\omega}\, \bigr) \,,
 \end{split}
\ee
where we have used the integral~$J$ given in~\eqref{eq:gfR2int}.
Further dropping~$e^{-\omega/T}$, we have
\be \label{eq:TlomquantA}
\begin{split}
\text{Im} \, \CG^{\Delta=1}_{_R} (\omega)  & \=
  \omega \,  \coth \bigl(\pi \sqrt{2 C\omega}\, \bigr) \\
& \= \begin{cases}
 \omega \bigl(1+ e^{- 2 \pi \sqrt{2 C\omega}}\bigr)  \,, \qquad & C\omega \gg 1 \,, \\
  \dfrac{\omega}{\sqrt{2\pi^2 C\omega}} \Bigl(1+\dfrac{2\pi^2}{3} \, C\omega + \dfrac{4\pi^4}{45} \, (C\omega)^2 + \dots  \Bigr) \,, \qquad & C\omega \ll 1 \,.
\end{cases}
\end{split}
\ee

\end{appendix}

\bibliography{NearExtremalAdSBH}

@article{Ge:2018lzo,
    author = "Ge, Xian-Hui and Jian, Shao-Kai and Wang, Yi-Li and Xian, Zhuo-Yu and Yao, Hong",
    title = "{Violation of the viscosity/entropy bound in translationally invariant non-Fermi liquids}",
    eprint = "1810.00669",
    archivePrefix = "arXiv",
    primaryClass = "hep-th",
    doi = "10.1103/PhysRevResearch.2.023366",
    journal = "Phys. Rev. Res.",
    volume = "2",
    number = "2",
    pages = "023366",
    year = "2020"
}

@article{Breitenlohner:1982jf,
    author = "Breitenlohner, Peter and Freedman, Daniel Z.",
    title = "{Stability in Gauged Extended Supergravity}",
    reportNumber = "Print-82-0500 (MIT)",
    doi = "10.1016/0003-4916(82)90116-6",
    journal = "Annals Phys.",
    volume = "144",
    pages = "249",
    year = "1982"
}

@article{Klebanov:1999tb,
    author = "Klebanov, Igor R. and Witten, Edward",
    title = "{AdS / CFT correspondence and symmetry breaking}",
    eprint = "hep-th/9905104",
    archivePrefix = "arXiv",
    reportNumber = "PUPT-1863, IASSNS-HEP-99-49",
    doi = "10.1016/S0550-3213(99)00387-9",
    journal = "Nucl. Phys. B",
    volume = "556",
    pages = "89--114",
    year = "1999"
}

@article{Witten:2001ua,
    author = "Witten, Edward",
    title = "{Multitrace operators, boundary conditions, and AdS / CFT correspondence}",
    eprint = "hep-th/0112258",
    archivePrefix = "arXiv",
    month = "12",
    year = "2001"
}

@article{Heemskerk:2010hk,
    author = "Heemskerk, Idse and Polchinski, Joseph",
    title = "{Holographic and Wilsonian Renormalization Groups}",
    eprint = "1010.1264",
    archivePrefix = "arXiv",
    primaryClass = "hep-th",
    doi = "10.1007/JHEP06(2011)031",
    journal = "JHEP",
    volume = "06",
    pages = "031",
    year = "2011"
}

@article{Faulkner:2010jy,
    author = "Faulkner, Thomas and Liu, Hong and Rangamani, Mukund",
    title = "{Integrating out geometry: Holographic Wilsonian RG and the membrane paradigm}",
    eprint = "1010.4036",
    archivePrefix = "arXiv",
    primaryClass = "hep-th",
    reportNumber = "MIT-CTP-4185, DCPT-10-47",
    doi = "10.1007/JHEP08(2011)051",
    journal = "JHEP",
    volume = "08",
    pages = "051",
    year = "2011"
}

@article{Gouteraux:2025exs,
    author = "Gout{\'e}raux, Blaise and Ramirez, David M. and Supiot, Cl{\'e}ment",
    title = {{Schwarzian quantum corrections to shear correlators of the near-extremal Reissner-Nordstr{\"o}m-AdS black hole}},
    eprint = "2512.19642",
    archivePrefix = "arXiv",
    primaryClass = "hep-th",
    reportNumber = "CPHT-R063.122025",
    month = "12",
    year = "2025"
}

@article{Sachdev:1992fk,
    author = "Sachdev, Subir and Ye, Jinwu",
    title = "{Gapless spin fluid ground state in a random, quantum Heisenberg magnet}",
    eprint = "cond-mat/9212030",
    archivePrefix = "arXiv",
    reportNumber = "PRINT-93-0077",
    doi = "10.1103/PhysRevLett.70.3339",
    journal = "Phys. Rev. Lett.",
    volume = "70",
    pages = "3339",
    year = "1993"
}

@article{Choi:2024xnv,
    author = "Choi, Sunjin and Jain, Diksha and Kim, Seok and Krishna, Vineeth and Lee, Eunwoo and Minwalla, Shiraz and Patel, Chintan",
    title = "{Dual dressed black holes as the end point of the charged superradiant instability in $\mathcal{N} = 4$ Yang Mills}",
    eprint = "2409.18178",
    archivePrefix = "arXiv",
    primaryClass = "hep-th",
    reportNumber = "TIFR/TH/24-19, LCTP-24-17",
    doi = "10.21468/SciPostPhys.18.4.137",
    journal = "SciPost Phys.",
    volume = "18",
    number = "4",
    pages = "137",
    year = "2025"
}

@article{Anninos:2011vn,
    author = "Anninos, Dionysios and Anous, Tarek and Barandes, Jacob and Denef, Frederik and Gaasbeek, Bram",
    title = "{Hot Halos and Galactic Glasses}",
    eprint = "1108.5821",
    archivePrefix = "arXiv",
    primaryClass = "hep-th",
    doi = "10.1007/JHEP01(2012)003",
    journal = "JHEP",
    volume = "01",
    pages = "003",
    year = "2012"
}

@article{Anninos:2010sq,
    author = "Anninos, Dionysios and Hartnoll, Sean A. and Iqbal, Nabil",
    title = "{Holography and the Coleman-Mermin-Wagner theorem}",
    eprint = "1005.1973",
    archivePrefix = "arXiv",
    primaryClass = "hep-th",
    doi = "10.1103/PhysRevD.82.066008",
    journal = "Phys. Rev. D",
    volume = "82",
    pages = "066008",
    year = "2010"
}

@article{Banerjee:2008th,
    author = "Banerjee, Nabamita and Bhattacharya, Jyotirmoy and Bhattacharyya, Sayantani and Dutta, Suvankar and Loganayagam, R. and Surowka, P.",
    title = "{Hydrodynamics from charged black branes}",
    eprint = "0809.2596",
    archivePrefix = "arXiv",
    primaryClass = "hep-th",
    doi = "10.1007/JHEP01(2011)094",
    journal = "JHEP",
    volume = "01",
    pages = "094",
    year = "2011"
}

@article{Almheiri:2014cka,
    author = "Almheiri, Ahmed and Polchinski, Joseph",
    title = "{Models of AdS$_{2}$ backreaction and holography}",
    eprint = "1402.6334",
    archivePrefix = "arXiv",
    primaryClass = "hep-th",
    doi = "10.1007/JHEP11(2015)014",
    journal = "JHEP",
    volume = "11",
    pages = "014",
    year = "2015"
}

@article{Hadar:2020kry,
    author = "Hadar, Shahar and Lupsasca, Alexandru and Porfyriadis, Achilleas P.",
    title = "{Extreme Black Hole Anabasis}",
    eprint = "2012.06562",
    archivePrefix = "arXiv",
    primaryClass = "hep-th",
    doi = "10.1007/JHEP03(2021)223",
    journal = "JHEP",
    volume = "03",
    pages = "223",
    year = "2021"
}

@article{Daguerre:2023cyx,
    author = "Daguerre, Lucas",
    title = "{Boundary correlators and the Schwarzian mode}",
    eprint = "2310.19885",
    archivePrefix = "arXiv",
    primaryClass = "hep-th",
    doi = "10.1007/JHEP01(2024)118",
    journal = "JHEP",
    volume = "01",
    pages = "118",
    year = "2024"
}

@article{Cremonini:2025yqe,
    author = "Cremonini, Sera and Li, Li and Liu, Xiao-Long and Nian, Jun",
    title = "{Quantum Corrections to $η/s$ from JT Gravity}",
    eprint = "2510.21602",
    archivePrefix = "arXiv",
    primaryClass = "hep-th",
    month = "10",
    year = "2025"
}

@article{Nian:2025oei,
    author = "Nian, Jun and Pando Zayas, Leopoldo A. and Yue, Cong-Yuan",
    title = "{Quantum Corrections in the Low-Temperature Fluid/Gravity Correspondence}",
    eprint = "2510.15411",
    archivePrefix = "arXiv",
    primaryClass = "hep-th",
    reportNumber = "PCFT-25-43, LITP-25-01",
    month = "10",
    year = "2025"
}

@article{PandoZayas:2025snm,
    author = "Pando Zayas, Leopoldo A. and Zhang, Jingchao",
    title = "{One-loop Corrected Holographic Shear Viscosity to Entropy Density Ratio at Low Temperatures}",
    eprint = "2510.16100",
    archivePrefix = "arXiv",
    primaryClass = "hep-th",
    reportNumber = "LITP-25-12",
    month = "10",
    year = "2025"
}

@article{Iqbal:2008by,
    author = "Iqbal, Nabil and Liu, Hong",
    title = "{Universality of the hydrodynamic limit in AdS/CFT and the membrane paradigm}",
    eprint = "0809.3808",
    archivePrefix = "arXiv",
    primaryClass = "hep-th",
    reportNumber = "MIT-CTP-3983",
    doi = "10.1103/PhysRevD.79.025023",
    journal = "Phys. Rev. D",
    volume = "79",
    pages = "025023",
    year = "2009"
}

@article{Betzios:2025sct,
    author = "Betzios, Panos and Papadoulaki, Olga and Zhou, Yanjun",
    title = "{Near-extremal quantum cross-section for charged fields and superradiance}",
    eprint = "2507.13896",
    archivePrefix = "arXiv",
    primaryClass = "hep-th",
    month = "7",
    year = "2025"
}

@article{Preskill:1991tb,
    author = "Preskill, John and Schwarz, Patricia and Shapere, Alfred D. and Trivedi, Sandip and Wilczek, Frank",
    title = "{Limitations on the statistical description of black holes}",
    reportNumber = "IASSNS-HEP-91-34, CALT-68-1730",
    doi = "10.1142/S0217732391002773",
    journal = "Mod. Phys. Lett. A",
    volume = "6",
    pages = "2353--2362",
    year = "1991"
}

@misc{kitaev,
    title = {Alexei {Kitaev}, {Caltech} \& {KITP}, {A} simple model of quantum holography},
    url = {https://online.kitp.ucsb.edu/online/entangled15/kitaev/},
    urldate = {2025-11-14},
}

@article{Maldacena:2016hyu,
    author = "Maldacena, Juan and Stanford, Douglas",
    title = "{Remarks on the Sachdev-Ye-Kitaev model}",
    eprint = "1604.07818",
    archivePrefix = "arXiv",
    primaryClass = "hep-th",
    doi = "10.1103/PhysRevD.94.106002",
    journal = "Phys. Rev. D",
    volume = "94",
    number = "10",
    pages = "106002",
    year = "2016"
}

@article{Shiraz,
    author = "Bhattacharyya, Sayantani and Hubeny, Veronika E and Minwalla, Shiraz and Rangamani, Mukund",
    title = "{Nonlinear Fluid Dynamics from Gravity}",
    eprint = "0712.2456",
    archivePrefix = "arXiv",
    primaryClass = "hep-th",
    reportNumber = "TIFR-TH-07-44, DCPT-07-73, NI07097",
    doi = "10.1088/1126-6708/2008/02/045",
    journal = "JHEP",
    volume = "02",
    pages = "045",
    year = "2008"
}

@article{erd,
    author = "Erdmenger, Johanna and Haack, Michael and Kaminski, Matthias and Yarom, Amos",
    title = "{Fluid dynamics of R-charged black holes}",
    eprint = "0809.2488",
    archivePrefix = "arXiv",
    primaryClass = "hep-th",
    reportNumber = "LMU-ASC-48-08, MPP-2008-116",
    doi = "10.1088/1126-6708/2009/01/055",
    journal = "JHEP",
    volume = "01",
    pages = "055",
    year = "2009"
}

@article{Kunduri:2007vf,
    author = "Kunduri, Hari K. and Lucietti, James and Reall, Harvey S.",
    title = "{Near-horizon symmetries of extremal black holes}",
    eprint = "0705.4214",
    archivePrefix = "arXiv",
    primaryClass = "hep-th",
    reportNumber = "DCPT-07-25",
    doi = "10.1088/0264-9381/24/16/012",
    journal = "Class. Quant. Grav.",
    volume = "24",
    pages = "4169--4190",
    year = "2007"
}

@article{DP,
    author = "Davison, Richard A. and Parnachev, Andrei",
    title = "{Hydrodynamics of cold holographic matter}",
    eprint = "1303.6334",
    archivePrefix = "arXiv",
    primaryClass = "hep-th",
    doi = "10.1007/JHEP06(2013)100",
    journal = "JHEP",
    volume = "06",
    pages = "100",
    year = "2013"
}

@article{EJL1,
    author = "Edalati, Mohammad and Jottar, Juan I. and Leigh, Robert G.",
    title = "{Transport Coefficients at Zero Temperature from Extremal Black Holes}",
    eprint = "0910.0645",
    archivePrefix = "arXiv",
    primaryClass = "hep-th",
    doi = "10.1007/JHEP01(2010)018",
    journal = "JHEP",
    volume = "01",
    pages = "018",
    year = "2010"
}

@article{EJL3,
    author = "Edalati, Mohammad and Jottar, Juan I. and Leigh, Robert G.",
    title = "{Holography and the sound of criticality}",
    eprint = "1005.4075",
    archivePrefix = "arXiv",
    primaryClass = "hep-th",
    doi = "10.1007/JHEP10(2010)058",
    journal = "JHEP",
    volume = "10",
    pages = "058",
    year = "2010"
}

@article{Blaise,
    author = "Arean, Daniel and Davison, Richard A. and Gout{\'e}raux, Blaise and Suzuki, Kenta",
    title = "{Hydrodynamic Diffusion and Its Breakdown near AdS2 Quantum Critical Points}",
    eprint = "2011.12301",
    archivePrefix = "arXiv",
    primaryClass = "hep-th",
    reportNumber = "CPHT-089.112020, IFT-UAM/CSIC-20-165",
    doi = "10.1103/PhysRevX.11.031024",
    journal = "Phys. Rev. X",
    volume = "11",
    number = "3",
    pages = "031024",
    year = "2021"
}

@article{Blaise2,
    author = "Gout{\'e}raux, Blaise and Ramirez, David M. and Sanchez-Garitaonandia, Mikel and Supiot, Cl{\'e}ment",
    title = "{Near-extremal holographic charge correlators}",
    eprint = "2506.11974",
    archivePrefix = "arXiv",
    primaryClass = "hep-th",
    month = "6",
    year = "2025"
}

@article{Gralla,
    author = "Gralla, Samuel E. and Ravishankar, Arun and Zimmerman, Peter",
    title = "{Semi-local Quantum Criticality and the Instability of Extremal Planar Horizons}",
    eprint = "1808.07053",
    archivePrefix = "arXiv",
    primaryClass = "hep-th",
    doi = "10.1007/JHEP12(2018)087",
    journal = "JHEP",
    volume = "12",
    pages = "087",
    year = "2018"
}

@article{Trivedi,
    author = "Moitra, Upamanyu and Sake, Sunil Kumar and Trivedi, Sandip P.",
    title = "{Near-Extremal Fluid Mechanics}",
    eprint = "2005.00016",
    archivePrefix = "arXiv",
    primaryClass = "hep-th",
    reportNumber = "TIFR/TH/20-12",
    doi = "10.1007/JHEP02(2021)021",
    journal = "JHEP",
    volume = "02",
    pages = "021",
    year = "2021"
}

@article{preau,
    author = {J{\"a}rvinen, M. and Kiritsis, E. and Nitti, F. and Pr{\'e}au, E.},
    title = "{Holographic neutrino transport in dense strongly-coupled matter}",
    eprint = "2306.00192",
    archivePrefix = "arXiv",
    primaryClass = "astro-ph.HE",
    reportNumber = "APCTP Pre2023 - 004, CCTP-2023-3, ITCP-2023/3",
    doi = "10.1007/JHEP11(2023)139",
    journal = "JHEP",
    volume = "11",
    pages = "139",
    year = "2023"
}

@article{Kurchan,
    author = "Facoetti, Davide and Biroli, Giulio and Kurchan, Jorge and Reichman, David R.",
    title = "{Classical Glasses, Black Holes, and Strange Quantum Liquids}",
    eprint = "1906.09228",
    archivePrefix = "arXiv",
    primaryClass = "hep-th",
    doi = "10.1103/PhysRevB.100.205108",
    journal = "Phys. Rev. B",
    volume = "100",
    number = "20",
    pages = "205108",
    year = "2019"
}

@article{Jensen:2016pah,
    author = "Jensen, Kristan",
    title = "{Chaos in AdS$_2$ Holography}",
    eprint = "1605.06098",
    archivePrefix = "arXiv",
    primaryClass = "hep-th",
    doi = "10.1103/PhysRevLett.117.111601",
    journal = "Phys. Rev. Lett.",
    volume = "117",
    number = "11",
    pages = "111601",
    year = "2016"
}

@article{Witten,
    author = "Adkins, Gregory S. and Nappi, Chiara R. and Witten, Edward",
    title = "{Static Properties of Nucleons in the Skyrme Model}",
    reportNumber = "PRINT-83-0493 (IAS,PRINCETON)",
    doi = "10.1016/0550-3213(83)90559-X",
    journal = "Nucl. Phys. B",
    volume = "228",
    pages = "552",
    year = "1983"
}

@article{Iliesiu:2020qvm,
    author = "Iliesiu, Luca V. and Turiaci, Gustavo J.",
    title = "{The statistical mechanics of near-extremal black holes}",
    eprint = "2003.02860",
    archivePrefix = "arXiv",
    primaryClass = "hep-th",
    doi = "10.1007/JHEP05(2021)145",
    journal = "JHEP",
    volume = "05",
    pages = "145",
    year = "2021"
}

@article{Iliesiu:2022onk,
    author = "Iliesiu, Luca V. and Murthy, Sameer and Turiaci, Gustavo J.",
    title = "{Revisiting the logarithmic corrections to the black hole entropy}",
    eprint = "2209.13608",
    archivePrefix = "arXiv",
    primaryClass = "hep-th",
    doi = "10.1007/JHEP07(2025)058",
    journal = "JHEP",
    volume = "07",
    pages = "058",
    year = "2025"
}

@article{Iliesiu:2022kny,
    author = "Iliesiu, Luca V. and Murthy, Sameer and Turiaci, Gustavo J.",
    title = "{Black hole microstate counting from the gravitational path integral}",
    eprint = "2209.13602",
    archivePrefix = "arXiv",
    primaryClass = "hep-th",
    doi = "10.1007/JHEP08(2025)152",
    journal = "JHEP",
    volume = "08",
    pages = "152",
    year = "2025"
}

@article{Banerjee:2010qc,
    author = "Banerjee, Shamik and Gupta, Rajesh Kumar and Sen, Ashoke",
    title = "{Logarithmic Corrections to Extremal Black Hole Entropy from Quantum Entropy Function}",
    eprint = "1005.3044",
    archivePrefix = "arXiv",
    primaryClass = "hep-th",
    doi = "10.1007/JHEP03(2011)147",
    journal = "JHEP",
    volume = "03",
    pages = "147",
    year = "2011"
}

@article{Sen:2012kpz, 
   author = "Sen, Ashoke",
    title = "{Logarithmic Corrections to N=2 Black Hole Entropy: An Infrared Window into the Microstates}",
    eprint = "1108.3842",
    archivePrefix = "arXiv",
    primaryClass = "hep-th",
    doi = "10.1007/s10714-012-1336-5",
    journal = "Gen. Rel. Grav.",
    volume = "44",
    number = "5",
    pages = "1207--1266",
    year = "2012"
}

@article{Murthy:2015yfa,
    author = "Murthy, Sameer and Reys, Valentin",
    title = "{Functional determinants, index theorems, and exact quantum black hole entropy}",
    eprint = "1504.01400",
    archivePrefix = "arXiv",
    primaryClass = "hep-th",
    doi = "10.1007/JHEP12(2015)028",
    journal = "JHEP",
    volume = "12",
    pages = "028",
    year = "2015"
}

@article{Dabholkar:2014ema,
    author = "Dabholkar, Atish and Gomes, Joao and Murthy, Sameer",
    title = "{Nonperturbative black hole entropy and Kloosterman sums}",
    eprint = "1404.0033",
    archivePrefix = "arXiv",
    primaryClass = "hep-th",
    doi = "10.1007/JHEP03(2015)074",
    journal = "JHEP",
    volume = "03",
    pages = "074",
    year = "2015"
}

@article{Dabholkar:2011ec,
    author = "Dabholkar, Atish and Gomes, Joao and Murthy, Sameer",
    title = "{Localization {\&} Exact Holography}",
    eprint = "1111.1161",
    archivePrefix = "arXiv",
    primaryClass = "hep-th",
    doi = "10.1007/JHEP04(2013)062",
    journal = "JHEP",
    volume = "04",
    pages = "062",
    year = "2013"
}

@article{Dabholkar:2010uh,
    author = "Dabholkar, Atish and Gomes, Joao and Murthy, Sameer",
    title = "{Quantum black holes, localization and the topological string}",
    eprint = "1012.0265",
    archivePrefix = "arXiv",
    primaryClass = "hep-th",
    doi = "10.1007/JHEP06(2011)019",
    journal = "JHEP",
    volume = "06",
    pages = "019",
    year = "2011"
}

@article{Faulkner:2009wj,
    author = "Faulkner, Thomas and Liu, Hong and McGreevy, John and Vegh, David",
    title = "{Emergent quantum criticality, Fermi surfaces, and AdS(2)}",
    eprint = "0907.2694",
    archivePrefix = "arXiv",
    primaryClass = "hep-th",
    reportNumber = "MIT-CTP-4050",
    doi = "10.1103/PhysRevD.83.125002",
    journal = "Phys. Rev. D",
    volume = "83",
    pages = "125002",
    year = "2011"
}

@article{Sachdev:2015efa,
    author = "Sachdev, Subir",
    title = "{Bekenstein-Hawking Entropy and Strange Metals}",
    eprint = "1506.05111",
    archivePrefix = "arXiv",
    primaryClass = "hep-th",
    doi = "10.1103/PhysRevX.5.041025",
    journal = "Phys. Rev. X",
    volume = "5",
    number = "4",
    pages = "041025",
    year = "2015"
}

@article{Almheiri:2016fws,
    author = "Almheiri, Ahmed and Kang, Byungwoo",
    title = "{Conformal Symmetry Breaking and Thermodynamics of Near-Extremal Black Holes}",
    eprint = "1606.04108",
    archivePrefix = "arXiv",
    primaryClass = "hep-th",
    doi = "10.1007/JHEP10(2016)052",
    journal = "JHEP",
    volume = "10",
    pages = "052",
    year = "2016"
}

@article{Heydeman:2020hhw,
    author = "Heydeman, Matthew and Iliesiu, Luca V. and Turiaci, Gustavo J. and Zhao, Wenli",
    title = "{The statistical mechanics of near-BPS black holes}",
    eprint = "2011.01953",
    archivePrefix = "arXiv",
    primaryClass = "hep-th",
    reportNumber = "PUPT-2621",
    doi = "10.1088/1751-8121/ac3be9",
    journal = "J. Phys. A",
    volume = "55",
    number = "1",
    pages = "014004",
    year = "2022"
}

@article{Nayak:2018qej,
    author = "Nayak, Pranjal and Shukla, Ashish and Soni, Ronak M. and Trivedi, Sandip P. and Vishal, V.",
    title = "{On the Dynamics of Near-Extremal Black Holes}",
    eprint = "1802.09547",
    archivePrefix = "arXiv",
    primaryClass = "hep-th",
    reportNumber = "TIFR/TH/17-35, TIFR-TH-17-35",
    doi = "10.1007/JHEP09(2018)048",
    journal = "JHEP",
    volume = "09",
    pages = "048",
    year = "2018"
}

@article{Moitra:2018jqs,
    author = "Moitra, Upamanyu and Trivedi, Sandip P. and Vishal, V.",
    title = "{Extremal and near-extremal black holes and near-CFT$_{1}$}",
    eprint = "1808.08239",
    archivePrefix = "arXiv",
    primaryClass = "hep-th",
    doi = "10.1007/JHEP07(2019)055",
    journal = "JHEP",
    volume = "07",
    pages = "055",
    year = "2019"
}

@article{Castro:2018ffi,
    author = "Castro, Alejandra and Larsen, Finn and Papadimitriou, Ioannis",
    title = "{5D rotating black holes and the nAdS$_{2}$/nCFT$_{1}$ correspondence}",
    eprint = "1807.06988",
    archivePrefix = "arXiv",
    primaryClass = "hep-th",
    reportNumber = "KIAS-P18074",
    doi = "10.1007/JHEP10(2018)042",
    journal = "JHEP",
    volume = "10",
    pages = "042",
    year = "2018"
}

@article{Stanford:2017thb,
    author = "Stanford, Douglas and Witten, Edward",
    title = "{Fermionic Localization of the Schwarzian Theory}",
    eprint = "1703.04612",
    archivePrefix = "arXiv",
    primaryClass = "hep-th",
    doi = "10.1007/JHEP10(2017)008",
    journal = "JHEP",
    volume = "10",
    pages = "008",
    year = "2017"
}

@article{Sachdev:2019bjn,
    author = "Sachdev, Subir",
    title = "{Universal low temperature theory of charged black holes with AdS$_2$ horizons}",
    eprint = "1902.04078",
    archivePrefix = "arXiv",
    primaryClass = "hep-th",
    doi = "10.1063/1.5092726",
    journal = "J. Math. Phys.",
    volume = "60",
    number = "5",
    pages = "052303",
    year = "2019"
}

@article{Camporesi:1995fb,
    author = "Camporesi, Roberto and Higuchi, Atsushi",
    title = "{On the Eigen functions of the Dirac operator on spheres and real hyperbolic spaces}",
    eprint = "gr-qc/9505009",
    archivePrefix = "arXiv",
    reportNumber = "BUTP-95-12",
    doi = "10.1016/0393-0440(95)00042-9",
    journal = "J. Geom. Phys.",
    volume = "20",
    pages = "1--18",
    year = "1996"
}

@article{Mertens:2017mtv,
    author = "Mertens, Thomas G. and Turiaci, Gustavo J. and Verlinde, Herman L.",
    title = "{Solving the Schwarzian via the Conformal Bootstrap}",
    eprint = "1705.08408",
    archivePrefix = "arXiv",
    primaryClass = "hep-th",
    doi = "10.1007/JHEP08(2017)136",
    journal = "JHEP",
    volume = "08",
    pages = "136",
    year = "2017"
}

@article{Lam:2018pvp,
    author = "Lam, Ho Tat and Mertens, Thomas G. and Turiaci, Gustavo J. and Verlinde, Herman",
    title = "{Shockwave S-matrix from Schwarzian Quantum Mechanics}",
    eprint = "1804.09834",
    archivePrefix = "arXiv",
    primaryClass = "hep-th",
    doi = "10.1007/JHEP11(2018)182",
    journal = "JHEP",
    volume = "11",
    pages = "182",
    year = "2018"
}

@article{Yang:2018gdb,
    author = "Yang, Zhenbin",
    title = "{The Quantum Gravity Dynamics of Near Extremal Black Holes}",
    eprint = "1809.08647",
    archivePrefix = "arXiv",
    primaryClass = "hep-th",
    doi = "10.1007/JHEP05(2019)205",
    journal = "JHEP",
    volume = "05",
    pages = "205",
    year = "2019"
}

@article{Maldacena:2016upp,
    author = "Maldacena, Juan and Stanford, Douglas and Yang, Zhenbin",
    title = "{Conformal symmetry and its breaking in two dimensional Nearly Anti-de-Sitter space}",
    eprint = "1606.01857",
    archivePrefix = "arXiv",
    primaryClass = "hep-th",
    doi = "10.1093/ptep/ptw124",
    journal = "PTEP",
    volume = "2016",
    number = "12",
    pages = "12C104",
    year = "2016"
}

@article{Emparan:2025qqf,
    author = "Emparan, Roberto and Trezzi, Stefano",
    title = "{Quantum transparency of near-extremal black holes}",
    eprint = "2507.03398",
    archivePrefix = "arXiv",
    primaryClass = "hep-th",
    doi = "10.1007/JHEP10(2025)023",
    journal = "JHEP",
    volume = "10",
    pages = "023",
    year = "2025"
}

@article{Emparan:2025sao,
    author = "Emparan, Roberto",
    title = "{Quantum cross-section of near-extremal black holes}",
    eprint = "2501.17470",
    archivePrefix = "arXiv",
    primaryClass = "hep-th",
    doi = "10.1007/JHEP04(2025)122",
    journal = "JHEP",
    volume = "04",
    pages = "122",
    year = "2025"
}

@article{Biggs:2025nzs,
    author = "Biggs, Anna",
    title = "{Following the state of an evaporating charged black hole into the quantum gravity regime}",
    eprint = "2503.02051",
    archivePrefix = "arXiv",
    primaryClass = "hep-th",
    month = "3",
    year = "2025"
}

@article{Sen:2008yk,
    author = "Sen, Ashoke",
    title = "{Entropy Function and AdS(2) / CFT(1) Correspondence}",
    eprint = "0805.0095",
    archivePrefix = "arXiv",
    primaryClass = "hep-th",
    doi = "10.1088/1126-6708/2008/11/075",
    journal = "JHEP",
    volume = "11",
    pages = "075",
    year = "2008"
}

@article{Engelsoy:2016xyb,
    author = {Engels{\"o}y, Julius and Mertens, Thomas G. and Verlinde, Herman},
    title = "{An investigation of AdS$_{2}$ backreaction and holography}",
    eprint = "1606.03438",
    archivePrefix = "arXiv",
    primaryClass = "hep-th",
    doi = "10.1007/JHEP07(2016)139",
    journal = "JHEP",
    volume = "07",
    pages = "139",
    year = "2016"
}

@article{Mertens:2022irh,
    author = "Mertens, Thomas G. and Turiaci, Gustavo J.",
    title = "{Solvable models of quantum black holes: a review on Jackiw{\textendash}Teitelboim gravity}",
    eprint = "2210.10846",
    archivePrefix = "arXiv",
    primaryClass = "hep-th",
    doi = "10.1007/s41114-023-00046-1",
    journal = "Living Rev. Rel.",
    volume = "26",
    number = "1",
    pages = "4",
    year = "2023"
}

@article{Brown:2024ajk,
    author = "Brown, Adam R. and Iliesiu, Luca V. and Penington, Geoff and Usatyuk, Mykhaylo",
    title = "{The evaporation of charged black holes}",
    eprint = "2411.03447",
    archivePrefix = "arXiv",
    primaryClass = "hep-th",
    month = "11",
    year = "2024"
}

@article{Maldacena:1998uz,
    author = "Maldacena, Juan Martin and Michelson, Jeremy and Strominger, Andrew",
    title = "{Anti-de Sitter fragmentation}",
    eprint = "hep-th/9812073",
    archivePrefix = "arXiv",
    reportNumber = "HUTP-98-A088, UCSBTH-98-8",
    doi = "10.1088/1126-6708/1999/02/011",
    journal = "JHEP",
    volume = "02",
    pages = "011",
    year = "1999"
}

@article{Edalati:2009bi,
    author = "Edalati, Mohammad and Jottar, Juan I. and Leigh, Robert G.",
    title = "{Transport Coefficients at Zero Temperature from Extremal Black Holes}",
    eprint = "0910.0645",
    archivePrefix = "arXiv",
    primaryClass = "hep-th",
    doi = "10.1007/JHEP01(2010)018",
    journal = "JHEP",
    volume = "01",
    pages = "018",
    year = "2010"
}

@article{Edalati:2010hk,
    author = "Edalati, Mohammad and Jottar, Juan I. and Leigh, Robert G.",
    title = "{Shear Modes, Criticality and Extremal Black Holes}",
    eprint = "1001.0779",
    archivePrefix = "arXiv",
    primaryClass = "hep-th",
    doi = "10.1007/JHEP04(2010)075",
    journal = "JHEP",
    volume = "04",
    pages = "075",
    year = "2010"
}

@article{Davison:2013bxa,
    author = "Davison, Richard A. and Parnachev, Andrei",
    title = "{Hydrodynamics of cold holographic matter}",
    eprint = "1303.6334",
    archivePrefix = "arXiv",
    primaryClass = "hep-th",
    doi = "10.1007/JHEP06(2013)100",
    journal = "JHEP",
    volume = "06",
    pages = "100",
    year = "2013"
}

@article{Kodama:2003kk,
    author = "Kodama, Hideo and Ishibashi, Akihiro",
    title = "{Master equations for perturbations of generalized static black holes with charge in higher dimensions}",
    eprint = "hep-th/0308128",
    archivePrefix = "arXiv",
    doi = "10.1143/PTP.111.29",
    journal = "Prog. Theor. Phys.",
    volume = "111",
    pages = "29--73",
    year = "2004"
}

@article{Chamblin:1999tk,
    author = "Chamblin, Andrew and Emparan, Roberto and Johnson, Clifford V. and Myers, Robert C.",
    title = "{Charged AdS black holes and catastrophic holography}",
    eprint = "hep-th/9902170",
    archivePrefix = "arXiv",
    reportNumber = "DAMTP-1999-29, EHU-FT-9902, UK-99-02, MCGILL-99-07",
    doi = "10.1103/PhysRevD.60.064018",
    journal = "Phys. Rev. D",
    volume = "60",
    pages = "064018",
    year = "1999"
}

@article{Policastro:2001yc,
    author = "Policastro, G. and Son, Dan T. and Starinets, Andrei O.",
    title = "{The Shear viscosity of strongly coupled N=4 supersymmetric Yang-Mills plasma}",
    eprint = "hep-th/0104066",
    archivePrefix = "arXiv",
    reportNumber = "NYU-TH-01-04-02, SNS-PH-01-05",
    doi = "10.1103/PhysRevLett.87.081601",
    journal = "Phys. Rev. Lett.",
    volume = "87",
    pages = "081601",
    year = "2001"
}

@article{Policastro:2002se,
    author = "Policastro, Giuseppe and Son, Dam T. and Starinets, Andrei O.",
    title = "{From AdS / CFT correspondence to hydrodynamics}",
    eprint = "hep-th/0205052",
    archivePrefix = "arXiv",
    reportNumber = "INT-PUB-02-32",
    doi = "10.1088/1126-6708/2002/09/043",
    journal = "JHEP",
    volume = "09",
    pages = "043",
    year = "2002"
}

@article{Son:2002sd,
    author = "Son, Dam T. and Starinets, Andrei O.",
    title = "{Minkowski space correlators in AdS / CFT correspondence: Recipe and applications}",
    eprint = "hep-th/0205051",
    archivePrefix = "arXiv",
    reportNumber = "INT-PUB-02-34",
    doi = "10.1088/1126-6708/2002/09/042",
    journal = "JHEP",
    volume = "09",
    pages = "042",
    year = "2002"
}

@article{Bagrets:2017pwq,
    author = "Bagrets, Dmitry and Altland, Alexander and Kamenev, Alex",
    title = "{Power-law out of time order correlation functions in the SYK model}",
    eprint = "1702.08902",
    archivePrefix = "arXiv",
    primaryClass = "cond-mat.str-el",
    doi = "10.1016/j.nuclphysb.2017.06.012",
    journal = "Nucl. Phys. B",
    volume = "921",
    pages = "727--752",
    year = "2017"
}

@article{Bagrets:2016cdf,
    author = "Bagrets, Dmitry and Altland, Alexander and Kamenev, Alex",
    editor = "Unno, Yoshinobu and Ohsugi, Takashi and Hou, Suen and Sadrozinski, Hartmut F. -W. and Lou, Xinchou and Zhu, Hongbo and Ouyang, Qun",
    title = "{Sachdev{\textendash}Ye{\textendash}Kitaev model as Liouville quantum mechanics}",
    eprint = "1607.00694",
    archivePrefix = "arXiv",
    primaryClass = "cond-mat.str-el",
    doi = "10.1016/j.nuclphysb.2016.08.002",
    journal = "Nucl. Phys. B",
    volume = "911",
    pages = "191--205",
    year = "2016"
}

@article{Aretakis:2011ha,
    author = "Aretakis, Stefanos",
    title = {{Stability and Instability of Extreme Reissner-Nordstr{\"o}m Black Hole Spacetimes for Linear Scalar Perturbations I}},
    eprint = "1110.2007",
    archivePrefix = "arXiv",
    primaryClass = "gr-qc",
    doi = "10.1007/s00220-011-1254-5",
    journal = "Commun. Math. Phys.",
    volume = "307",
    pages = "17--63",
    year = "2011"
}

@article{Aretakis:2012ei,
    author = "Aretakis, Stefanos",
    title = "{Horizon Instability of Extremal Black Holes}",
    eprint = "1206.6598",
    archivePrefix = "arXiv",
    primaryClass = "gr-qc",
    doi = "10.4310/ATMP.2015.v19.n3.a1",
    journal = "Adv. Theor. Math. Phys.",
    volume = "19",
    pages = "507--530",
    year = "2015"
}

@article{Lucietti:2012xr,
    author = "Lucietti, James and Murata, Keiju and Reall, Harvey S. and Tanahashi, Norihiro",
    title = {{On the horizon instability of an extreme Reissner-Nordstr{\"o}m black hole}},
    eprint = "1212.2557",
    archivePrefix = "arXiv",
    primaryClass = "gr-qc",
    doi = "10.1007/JHEP03(2013)035",
    journal = "JHEP",
    volume = "03",
    pages = "035",
    year = "2013"
}

@article{Lucietti:2012sf,
    author = "Lucietti, James and Reall, Harvey S.",
    title = "{Gravitational instability of an extreme Kerr black hole}",
    eprint = "1208.1437",
    archivePrefix = "arXiv",
    primaryClass = "gr-qc",
    doi = "10.1103/PhysRevD.86.104030",
    journal = "Phys. Rev. D",
    volume = "86",
    pages = "104030",
    year = "2012"
}

@article{Murata:2012ct,
    author = "Murata, Keiju",
    title = "{Instability of higher dimensional extreme black holes}",
    eprint = "1211.6903",
    archivePrefix = "arXiv",
    primaryClass = "gr-qc",
    doi = "10.1088/0264-9381/30/7/075002",
    journal = "Class. Quant. Grav.",
    volume = "30",
    pages = "075002",
    year = "2013"
}

@article{Hadar:2018izi,
    author = "Hadar, Shahar",
    title = "{Near-extremal black holes at late times, backreacted}",
    eprint = "1811.01022",
    archivePrefix = "arXiv",
    primaryClass = "hep-th",
    doi = "10.1007/JHEP01(2019)214",
    journal = "JHEP",
    volume = "01",
    pages = "214",
    year = "2019"
}

@article{KU,
    author = "Kehle, Christoph and Unger, Ryan",
    title = "{Gravitational collapse to extremal black holes and the third law of black hole thermodynamics}",
    eprint = "2211.15742",
    archivePrefix = "arXiv",
    primaryClass = "gr-qc",
    month = "11",
    year = "2022"
}
\bibliographystyle{JHEP}

\end{document}